\documentclass[10pt,twocolumn]{article}
\usepackage[top=1.8cm, bottom=1.9cm, left=1.7cm, right=1.7cm]{geometry}
\usepackage{helvet}  %
\usepackage{courier}  %
\usepackage[hyphens]{url}  %
\usepackage{graphicx}  %
\usepackage{url}

\usepackage{xcolor}
\usepackage{booktabs}
\usepackage{graphicx}
\usepackage{paralist}
\usepackage[]{caption}
\usepackage{subfigure}
\usepackage{stackengine}
\usepackage{times}
\usepackage{pifont}
\usepackage[hang,flushmargin]{footmisc}

\usepackage[keeplastbox]{flushend}

\usepackage[bookmarks=true,%
            bookmarksnumbered=true,%
            colorlinks=true,%
            linkcolor=linkcol,%
            citecolor=citecol,%
            urlcolor=urlcol,%
            hypertexnames=true,%
            pdfpagelabels]%
      {hyperref}

\definecolor{linkcol}{rgb}{0.3,0,0}
\definecolor{citecol}{rgb}{0.3,0,0}
\definecolor{urlcol}{rgb}{0.3,0,0}

\usepackage{xspace}
\usepackage{comment}

\usepackage{todonotes} %

\newcommand{\descr}[1]{\smallskip\noindent\textbf{#1}}

\makeatletter
\def\url@leostyle{%
  \@ifundefined{selectfont}{\def\UrlFont{}}%
  {\def\UrlFont{}}%
}
\makeatother
\usepackage[hyphenbreaks]{breakurl}

\newcommand{\dspol}{{/pol/}\xspace}

\newcommand{\wvcontrol}[0]{$\mathcal{W}_C$\xspace{}}
\newcommand{\wvall}[0]{$\mathcal{W}_A$\xspace{}}
\newcommand{\wvweekly}[0]{$\mathcal{W}_{t=i}~,i\in{\mathcal{T}}$\xspace{}}
\newcommand{\duration}[0]{$\mathcal{T}$\xspace{}}

\begin{document}

\title{\bf{``Go eat a bat, Chang!'': On the Emergence of Sinophobic Behavior on Web Communities in the Face of COVID-19}\thanks{This is the full version of the paper, with same title, appearing in the Proceedings of the 30th The Web Conference (WWW 2021). Please cite the WWW version.}}

\author{Fatemeh Tahmasbi$^1$, Leonard Schild$^2$, Chen Ling$^{3}$, Jeremy Blackburn$^1$, \\Gianluca Stringhini$^3$, Yang Zhang$^2$, and Savvas Zannettou$^{4}$\\[0.5ex]
$^1$Binghamton University,$^{2}$CISPA Helmholtz Center for Information Security,\\ $^3$Boston University,
$^4$Max Planck Institute for Informatics\\
\normalsize ftahmas1@binghamton.edu, leonard.schild@cispa.de, ccling@bu.edu, jblackbu@binghamton.edu, \\ \normalsize gian@bu.edu, zhang@cispa.de, szannett@mpi-inf.mpg.de\vspace*{-0.3cm}}\date{}

\maketitle

\begin{abstract}
The outbreak of the COVID-19 pandemic has changed our lives in an unprecedented way. 
In the face of the projected catastrophic consequences, many countries enforced social distancing measures in an attempt to limit the spread of the virus.
Under these conditions, the Web has become an indispensable medium for information acquisition, communication, and entertainment.
At the same time, unfortunately, the Web is exploited for the dissemination of potentially harmful and disturbing content, such as the spread of conspiracy theories and hateful speech towards specific ethnic groups, in particular towards Chinese people since COVID-19 is believed to have originated from China.

In this paper, we make a first attempt to study the emergence of Sinophobic behavior on the Web during the outbreak of the COVID-19 pandemic.
We collect two large-scale datasets from Twitter and 4chan's Politically Incorrect board (\dspol) over a time period of approximately five months and analyze them to investigate whether there is a rise or important differences with regard to the dissemination of Sinophobic content.
We find that COVID-19 indeed drives the rise of Sinophobia on the Web and the dissemination of Sinophobic content is a cross-platform phenomenon: it exists both on fringe Web communities, as well as mainstream ones like Twitter.
Also, using word embeddings over time, we characterize the evolution and emergence of new Sinophobic slurs on both Twitter and \dspol.
Finally, we find interesting differences in the context in which words related to Chinese people are used on the Web before and after the COVID-19 outbreak: on Twitter we observe a shift towards blaming China for the situation, while on \dspol we find a shift towards using more (and new) Sinophobic slurs.
\end{abstract}

\section{Introduction}

The coronavirus disease (COVID-19) caused by the severe acute respiratory syndrome coronavirus 2 (SARS-CoV-2) is the largest pandemic event of the information age.
SARS-CoV-2 is thought to have originated in China, with the presumed ground zero centered around a wet market in the city of Wuhan in the Hubei province~\cite{coronavirus_pandemic}.
In a few months, SARS-CoV-2 has spread, allegedly from a bat or pangolin, to essentially every country in the world, resulting in over 1M cases of COVID-19 and 50K deaths as of April 2, 2020~\cite{coronavirus_numbers}.

Humankind has taken unprecedented steps to mitigating the spread of SARS-CoV-2, enacting social distancing measures that go against our very nature.
While the repercussions of social distancing measures are yet to be fully understood, one thing is certain: the Web has not only proven essential to the approximately normal continuation of daily life, but also as a tool by which to ease the pain of isolation.

Unfortunately, just like the spread of COVID-19 was accelerated in part by international travel enabled by modern technology, the connected nature of the Web has enabled the spread of misinformation~\cite{batsoup_misinformation}, conspiracy theories~\cite{coronavirus_conspiracies}, and racist rhetoric~\cite{coronavirus_hate}.
Considering society's recent struggles with online racism (often leading to violence), and the politically charged nature of the SARS-CoV-2's emergence, there is every reason to believe that a wave of Sinophobia is not just coming, but already upon us.

In this paper, we present an analysis of how online Sinophobia has emerged, and evolved as the COVID-19 crisis has unfolded.
To do this, we collect and analyze two large-scale datasets obtained from Twitter and 4chan's Politically Incorrect board (\dspol).
Using temporal analysis, word embeddings, and graph analysis, we shed light into how prevalent is Sinophobic behavior on these communities, how this prevalence changes over time as the COVID-19 pandemic unfolds, and more importantly, we investigate whether there are substantial differences in discussions related to Chinese people by comparing the behavior pre-  and post- COVID-19 crisis.

\descr{Main findings.} Among others, we make the following findings:
\begin{compactenum}
\item We find a rise in discussions related to China and Chinese people on Twitter and 4chan's \dspol after the outbreak of the COVID-19 pandemic. 
At the same time, we observe a rise in the use of specific Sinophobic slurs on both Twitter and \dspol.
Also, by comparing our findings to real-world events, we find that the increase in these discussions and Sinophobic slurs coincides with real-world events related to the outbreak of the COVID-19 pandemic.
\item We find important differences with regard to the use of Sinophobic slurs across Twitter and \dspol. For instance, we find that the most popular Sinophobic slur on Twitter is ``chinazi,'' while on \dspol is ``chink.''
Furthermore, using word embeddings, we looked into the context of words used in discussions referencing Chinese people finding that various racial slurs are used in these contexts on both Twitter and \dspol.
This indicates that Sinophobic behavior is a cross-platform phenomenon existing in both fringe Web communities like \dspol and mainstream ones like Twitter.
\item Using word embeddings over time, we discover new emerging slurs and terms related to Sinophobic behavior, as well as the COVID-19 pandemic. 
For instance, on \dspol we observe the emergence of the term ``kungflu'' after January, 2020, while on Twitter we observe the emergence of the term ``asshoe,'' which aims to make fun of the accent of Chinese people speaking English.
\item By comparing our dataset pre- and post-COVID-19 outbreak, we observe shifts in the content posted by users on Twitter and \dspol. On Twitter, we observe a shift towards blaming China and Chinese people about the outbreak, while on \dspol we observe a shift towards using more, and new, Sinophobic slurs.
\end{compactenum}

\descr{Disclaimer.} Note that content posted on the Web communities we study is likely to be considered as highly offensive or racist. Throughout the rest of this paper, we do not censor any language, thus we warn the readers that content presented is likely to be offensive and upsetting.

\section{Related Works}

Due to its incredible impact to everybody's life in early 2020, the COVID-19 pandemic has already attracted the attention of researchers.
In particular, a number of papers studied how users on social media discussed this emergency.
Chen et al.~\cite{chen2020covid} release a dataset of 50M tweets releated to the pandemic.
Cinelli et al.~\cite{cinelli2020covid}, Singh et al.~\cite{singh2020first}, and Kouzy et al.~\cite{kouzy2020coronavirus} studied misinformation narratives about COVID-19 on Twitter.
Lopez et al.~\cite{lopez2020understanding} analyzed a multi-language Twitter dataset to understand how people in different countries reacted to policies related to COVID-19.

A number of papers studied racist activity on social networks.
Keum and Miller~\cite{keum2018racism} argued that racism on the Internet is pervasive and that users are likely to encounter it.
Zimmerman et al.~\cite{zimmerman2016online} focused on the influence that the anonymity brought by the Internet has on the likelihood for people to take part in online aggression.
Relia et al.~\cite{relia2019race} found that racist online activity correlates with hate crimes. 
In other words, users located in areas with higher occurrence of hate crimes are more likely to engage in racism on social media.
Yang and Counts~\cite{yang2018understanding} studied how users who experienced racism on Reddit self-narrate their experience.
They characterize the different types of racism experienced by users with different demographics, and show that commiseration is the most valued form of social support.

Zannettou et al.~\cite{zannettou2020quantitative} present a quantitative approach to understand racism targeting Jewish people online.
As part of their analysis, they present a method to quantify the evolution of racist language based on word embeddings, similar to the technique presented in this paper.
Hasanuzzaman et al.~\cite{hasanuzzaman2017demographic} investigated how demographic traits of Twitter users can act as a predictor of racist activity.
By modeling demographic traits as vector embeddings, they found that male and younger users (under 35) are more likely to engage in racism on Twitter.

Other work performed quantitative studies to characterize hateful users on social media, analyzing their language and their sentiment~\cite{chatzakou2017measuring,ribeiro2018characterizing}.
In particular, it focused on discrimination and hate directed against women, for example as part of the Pizzagate conspiracy~\cite{chatzakou2017hate,chess2015conspiracy}.

\descr{Remarks.} To the best of our knowledge, ours is the first data-driven study on the evolution of racist rhetoric against Chinese people and people of Asian descent in light of the COVID-19 pandemic.

\begin{table*}[t!]
	\centering
		\begin{tabular}{c | c | c}
		\toprule
		Number & Day & Event \\
		\midrule
		1 & December 12, 2019 & President Donald Trump signs an initial trade deal with China~\cite{tradewar_deal}. \\ 
		2 & January 23, 2020 & The Chinese government announces a lock-down of Wuhan and other cities in Hubei~\cite{wuhan_lockdown}.\\
		3 & January 30, 2020 & The World Health Organization declares a public health emergency~\cite{who_emergency}. \\
		4 & February 23, 2020 & 11 municipalities in Lombardy, Italy are locked down~\cite{lombardy_lockdown}. \\
		5 & March 9, 2020 & Italy extends restrictions in the northern region of the country~\cite{italy_lockdown}.\\
		6 & March 16, 2020 & Donald Trump referred to COVID-19 as ``Chinese Virus'' on Twitter~\cite{trump_chinesevirus}.\\
		\bottomrule
	\end{tabular}
	\caption{Major events, annotated on Figures~\ref{fig:temporal_china_4chan}--~\ref{fig:temporal_slurs_twitter}.}
	\label{tab:events}
\end{table*}

\section{Datasets}
To study the extent and evolution of Sinophobic behavior on the Web, we collect and analyze two large-scale datasets from Twitter and 4chan's Politically Incorrect board (\dspol).

\descr{Twitter.} Twitter is a popular mainstream microblog used by millions of users for disseminating information. 
To obtain data from Twitter, we leverage the Streaming API\footnote{\url{https://developer.twitter.com/en/docs/labs/sampled-stream/overview}}, which provides a 1\% random sample of all tweets made available on the platform.
We collect tweets posted between 28 October, 2019 and 22 March 2020, and then we filter only the ones posted in English, ultimately collecting 222,212,841 tweets.

\descr{4chan's \dspol.} 4chan is an imageboard that allows the anonymous posting of information. 
The imageboard is divided into several sub-communities called \emph{boards}: each board has its own topic of interest and moderation policy.
In this work, we focus on the Politically Incorrect board (\dspol), simply because it is the main board for the discussion of world events.
To collect data, we use the data collection approach from Hine et al.~\cite{hine2017kek}, to collect all posts made on \dspol between 28 October, 2019 and 22 March, 2020.
Overall, we collect 16,808,191 posts.

\descr{Remarks.} We elect to focus on these two specific Web communities, as we believe that they are representative examples of both mainstream and fringe Web communities.
That is, Twitter is a popular mainstream community that is used by hundreds of millions of users around the globe, while 4chan's \dspol is a notorious fringe Web community that is known for the dissemination of hateful or weaponized information~\cite{hine2017kek}.

\begin{figure*}[t!]
\center
\subfigure[]{\includegraphics[width=\columnwidth]{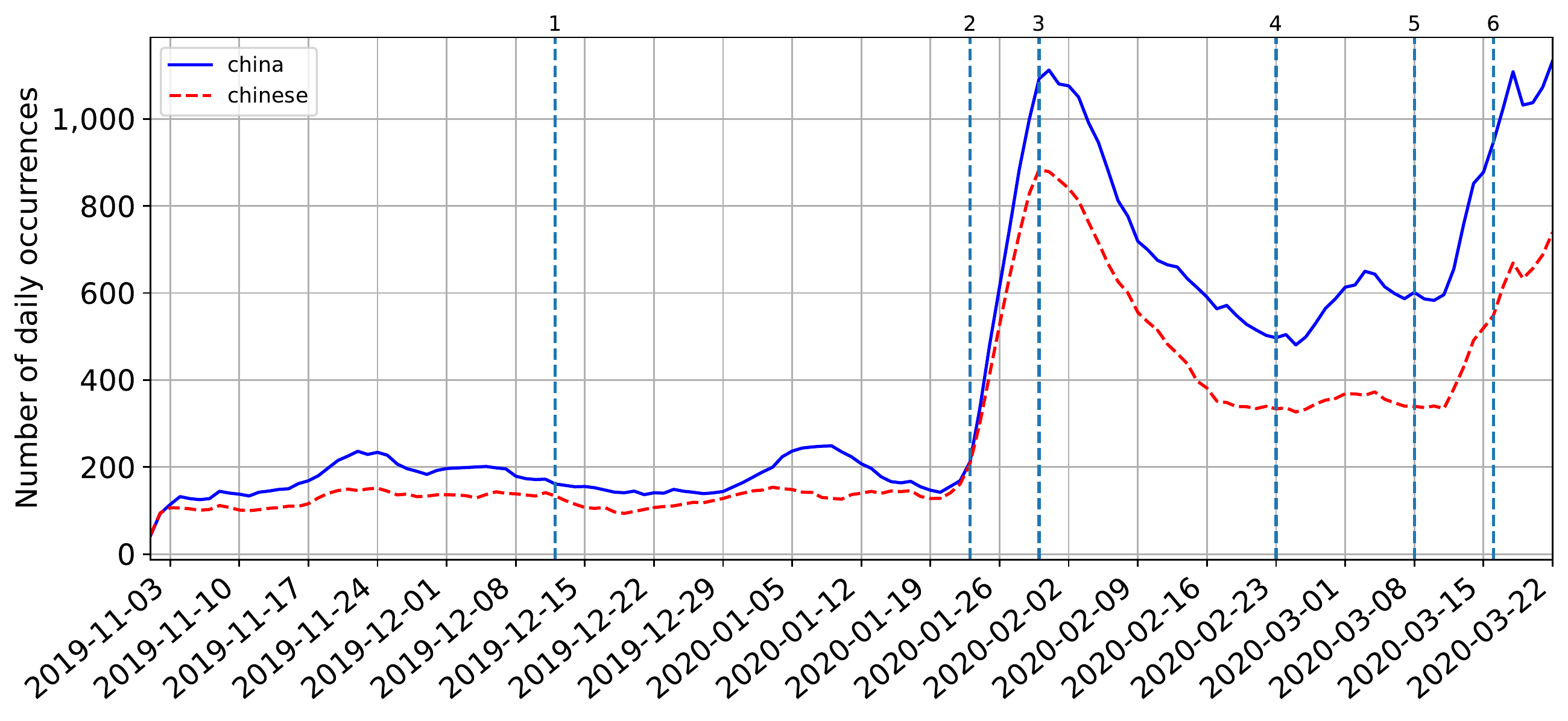}\label{fig:china_chinese_4chan_abs}}
\subfigure[]{\includegraphics[width=\columnwidth]{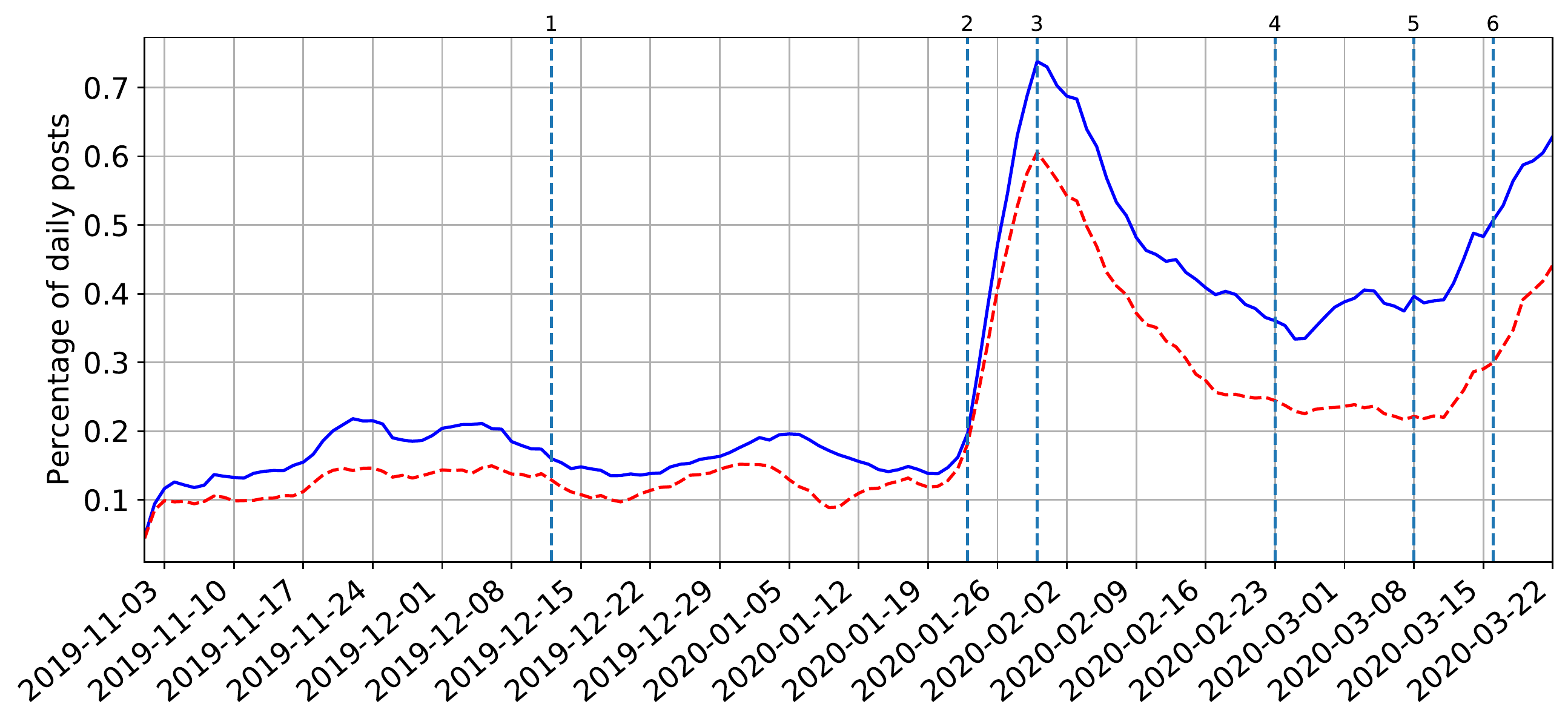}\label{fig:china_chinese_4chan_rel}}
\caption{Mentions of the terms ``china'' and ``chinese'' on 4chan's \dspol.} 
\label{fig:temporal_china_4chan}
\end{figure*}

\begin{figure*}[t!]
\center
\subfigure[]{\includegraphics[width=\columnwidth]{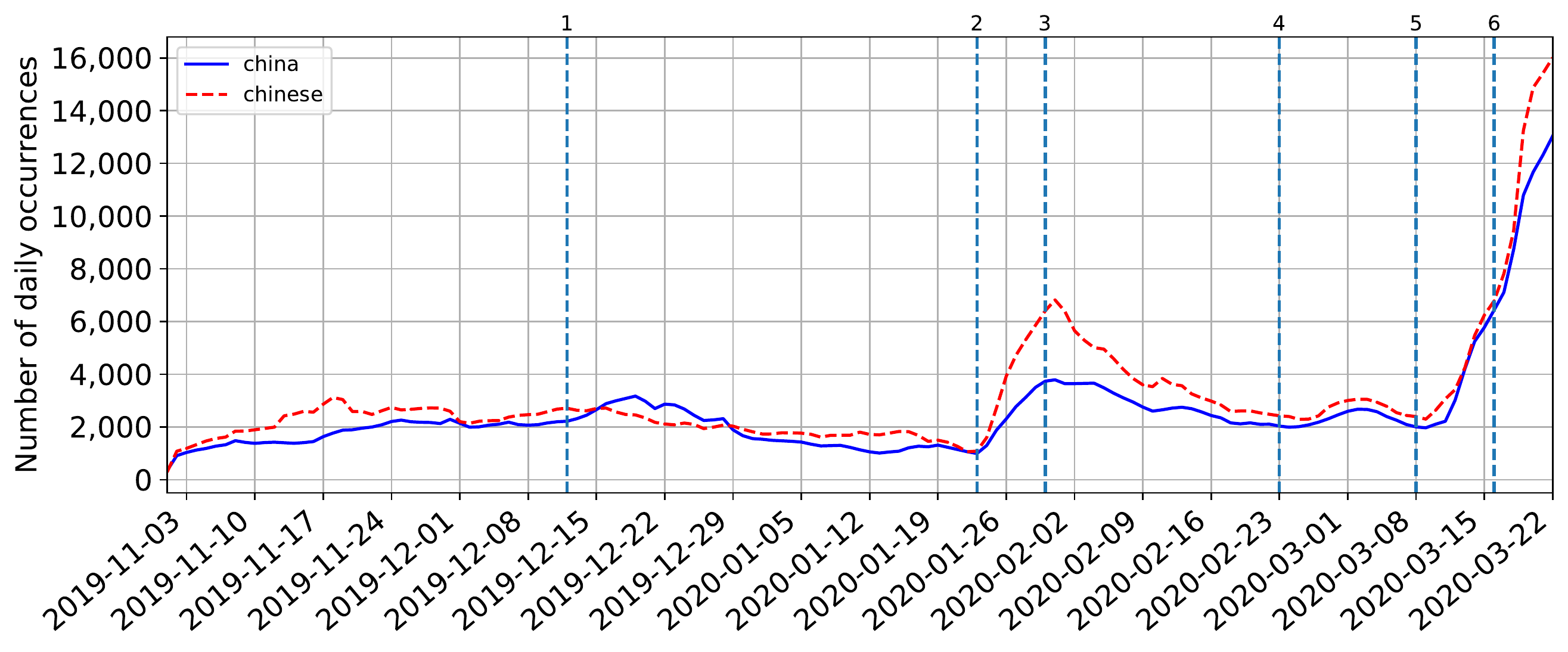}\label{fig:china_chinese_twitter_abs}}
\subfigure[]{\includegraphics[width=\columnwidth]{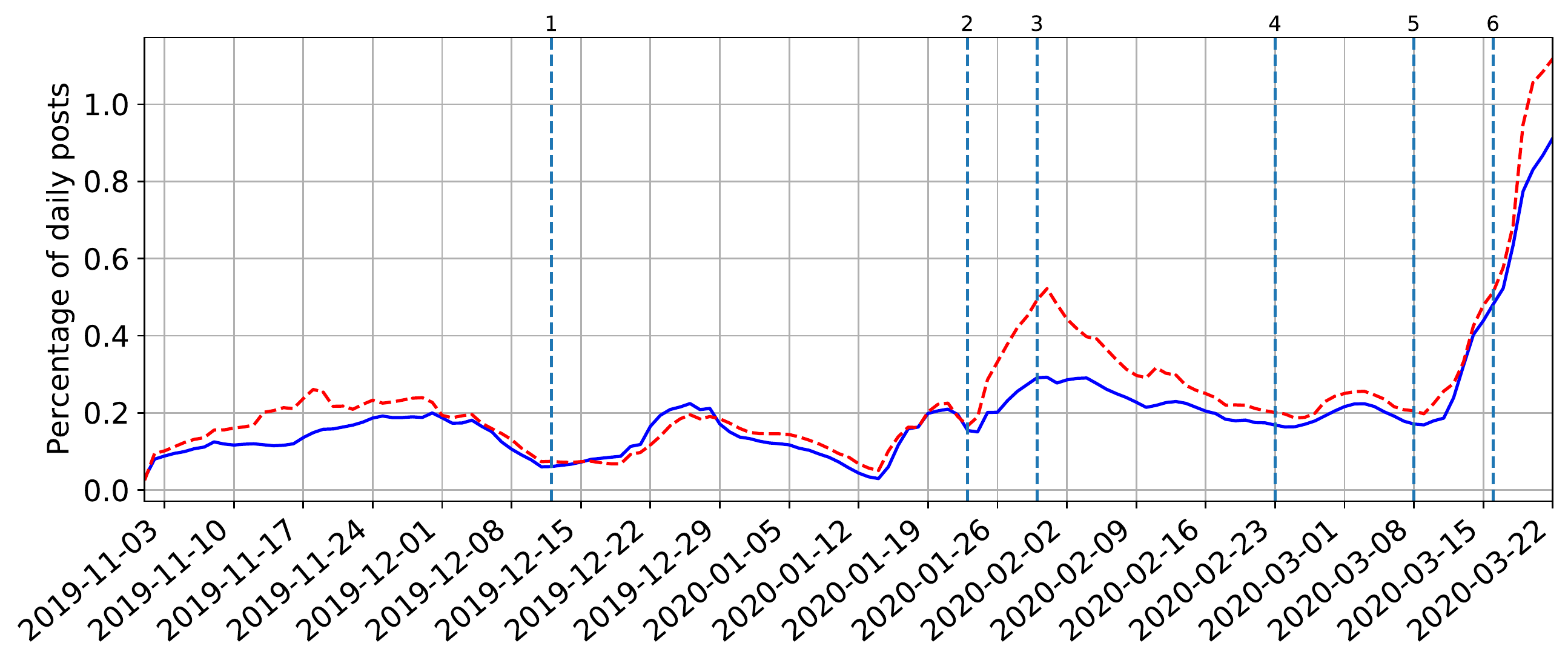}\label{fig:china_chinese_twitter_rel}}
\caption{Mentions of the terms ``china'' and ``chinese'' on Twitter.} 
\label{fig:temporal_china_twitter}
\end{figure*}

\begin{figure*}[t!]
\center
\subfigure[]{\includegraphics[width=\columnwidth]{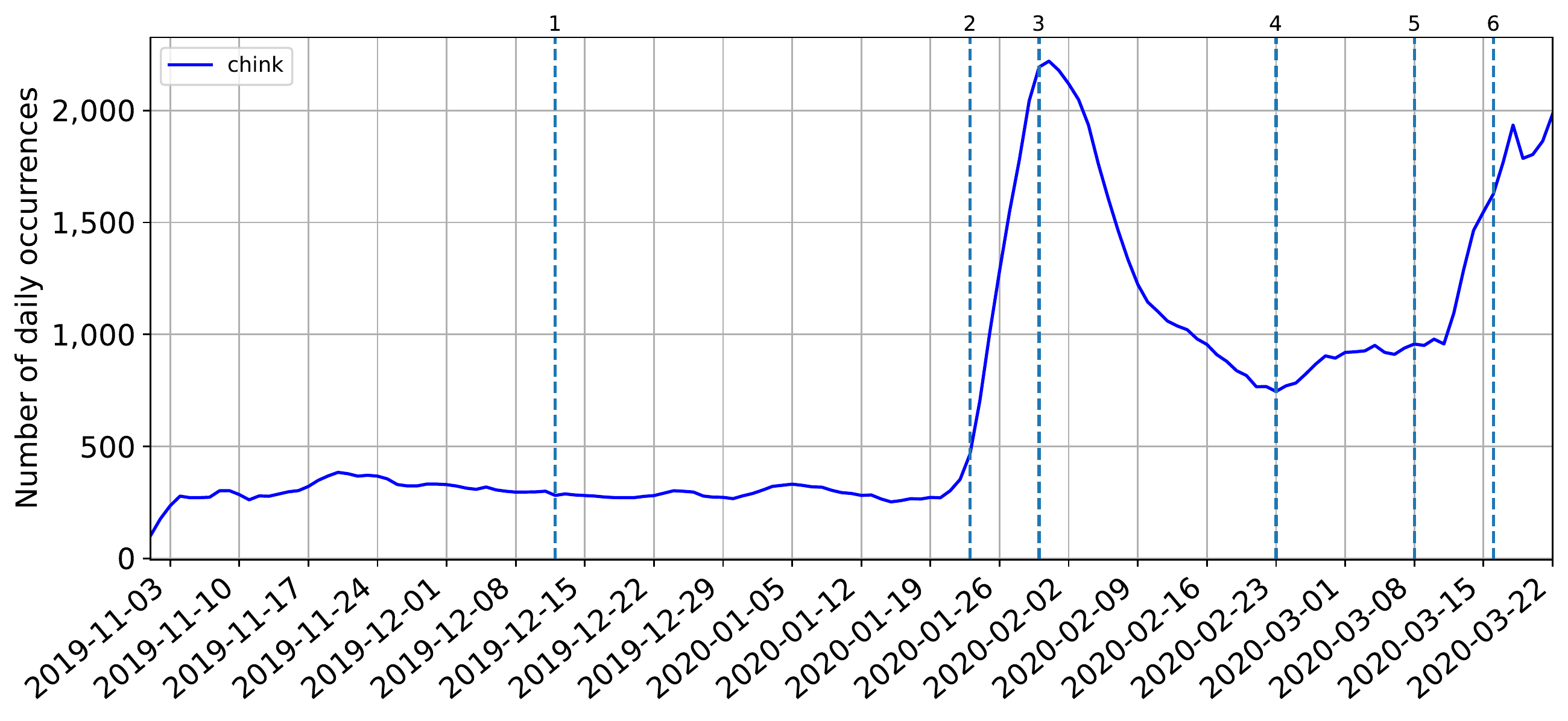}\label{fig:4chan_counts_chink_abs}}
\subfigure[]{\includegraphics[width=\columnwidth]{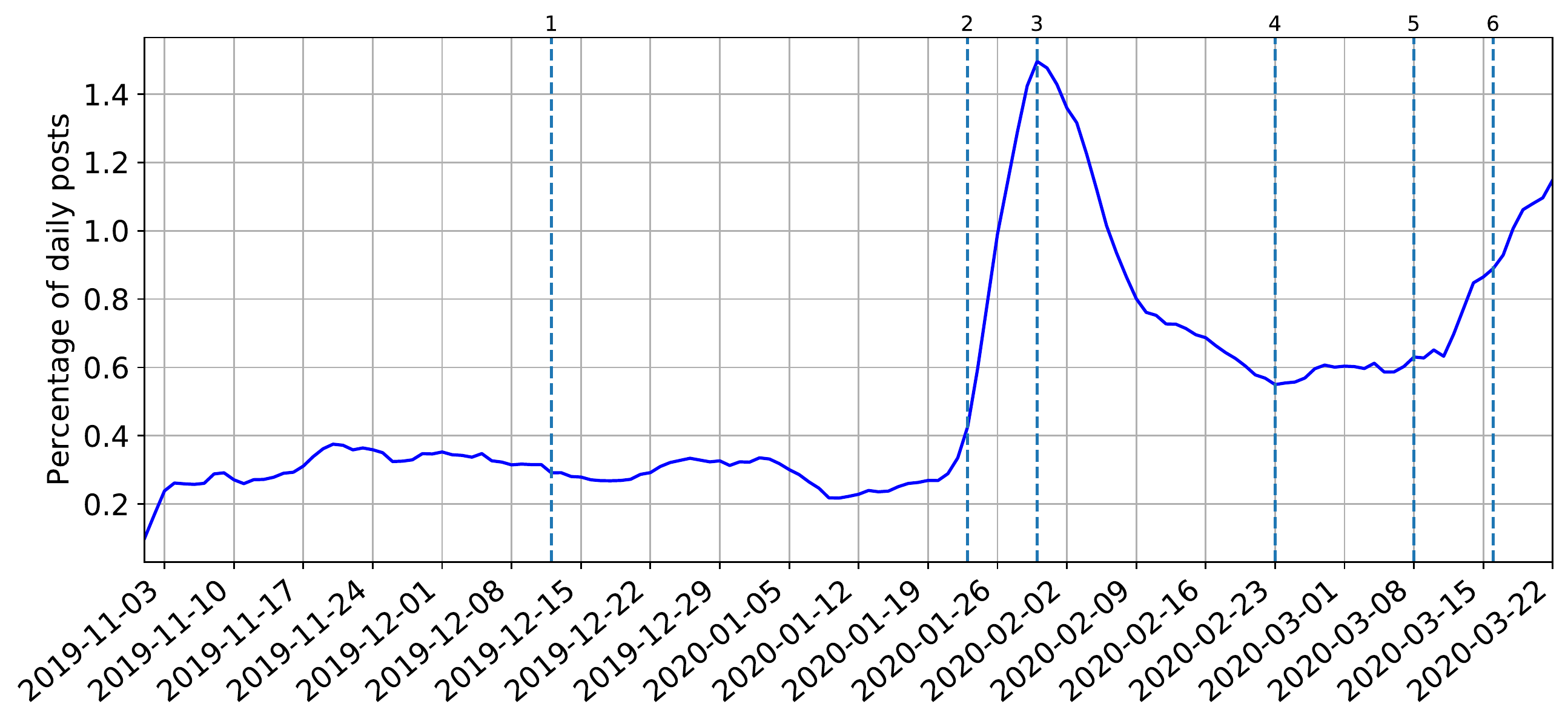}\label{fig:4chan_counts_chink_rel}}
\subfigure[]{\includegraphics[width=\columnwidth]{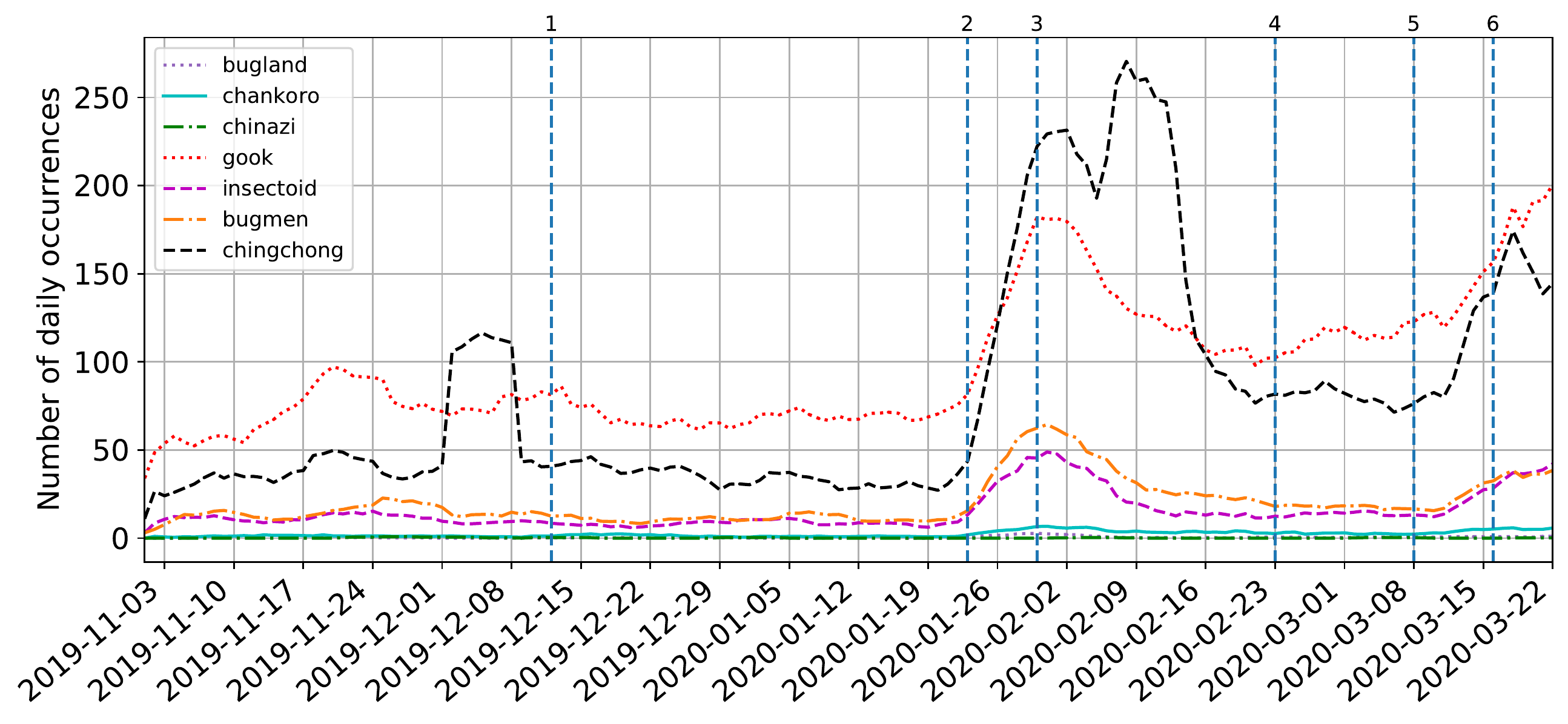}\label{fig:4chan_counts_other-slurs_abs}}
\subfigure[]{\includegraphics[width=\columnwidth]{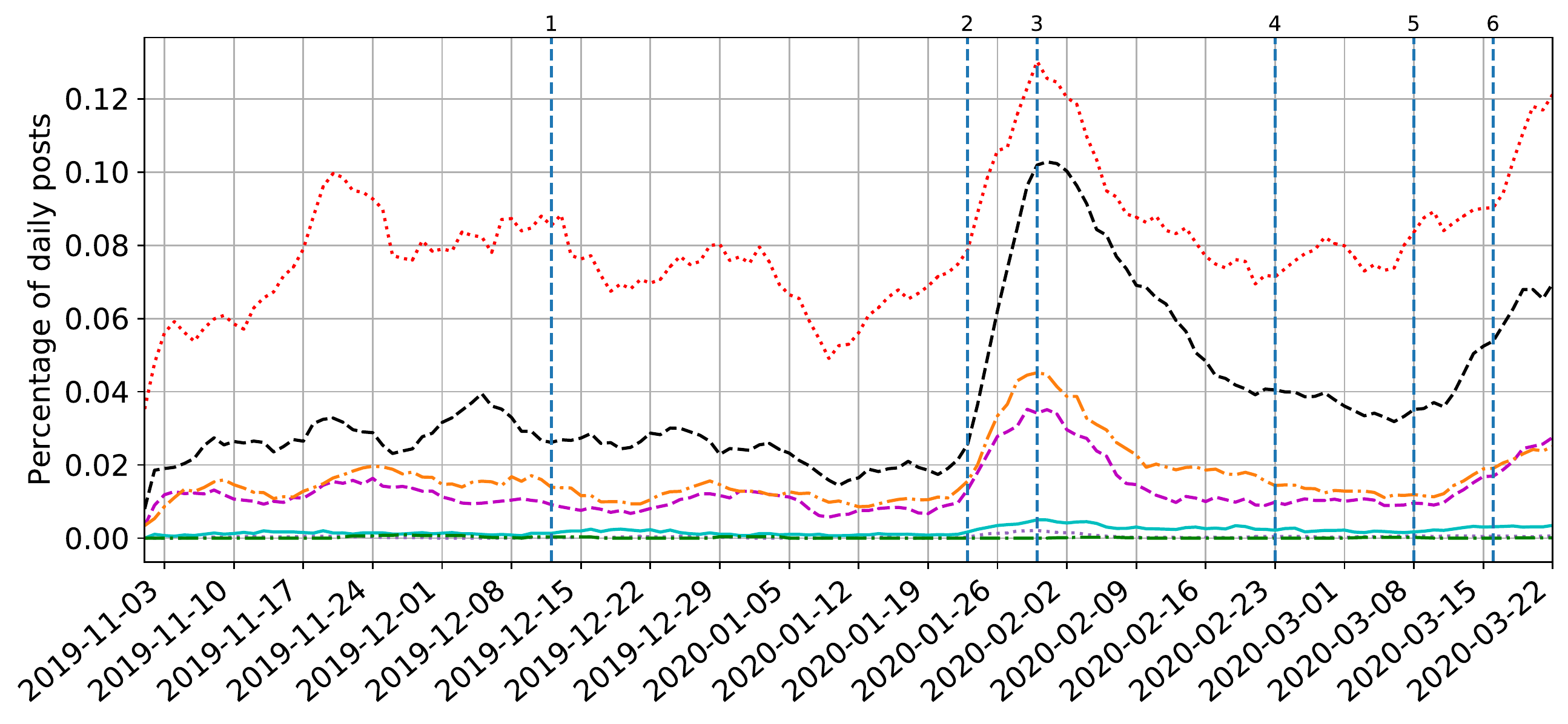}\label{fig:4chan_counts_other-slurs_rel}}
\caption{Mentions of Sinophobic racial slurs on 4chan's \dspol. }
\label{fig:temporal_slurs_4chan}
\end{figure*}

\begin{figure*}[t!]
\center
\subfigure[]{\includegraphics[width=\columnwidth]{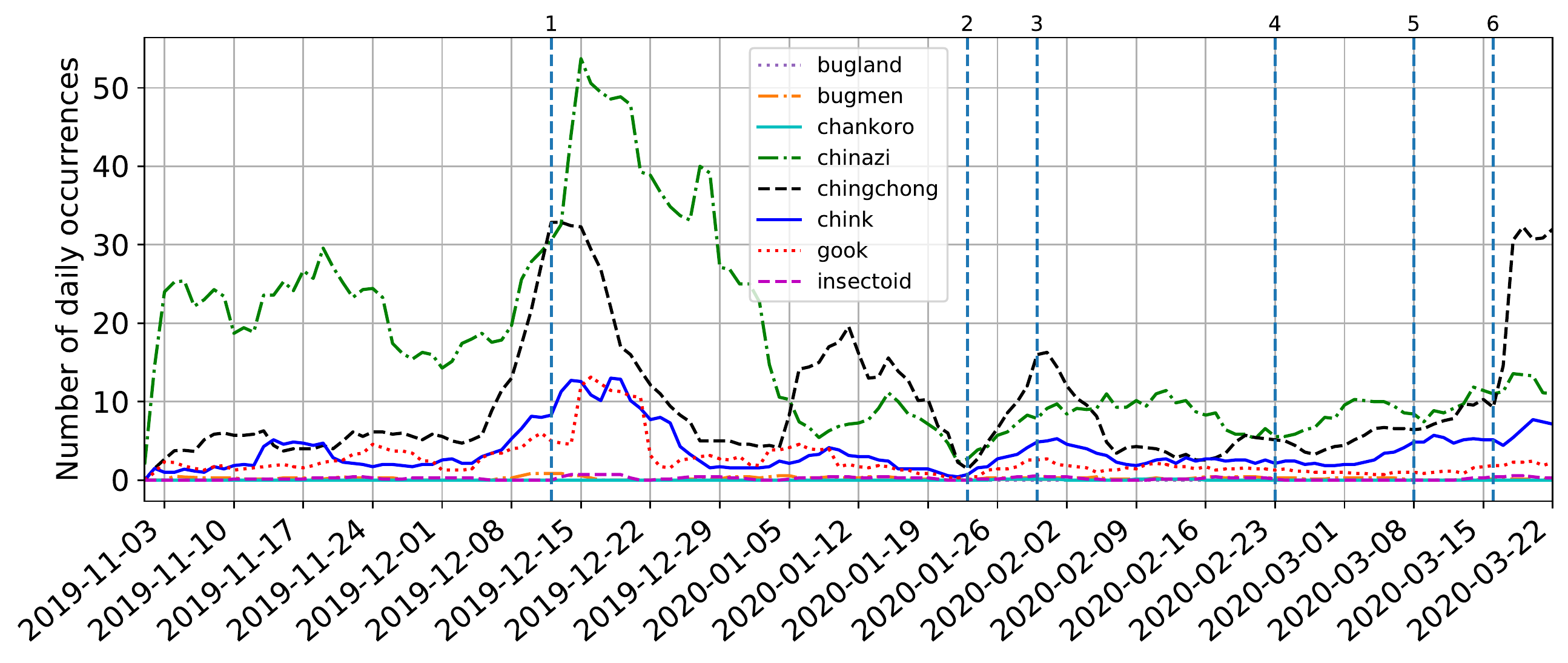}\label{fig:twitter_counts_other-slurs_abs}}
\subfigure[]{\includegraphics[width=\columnwidth]{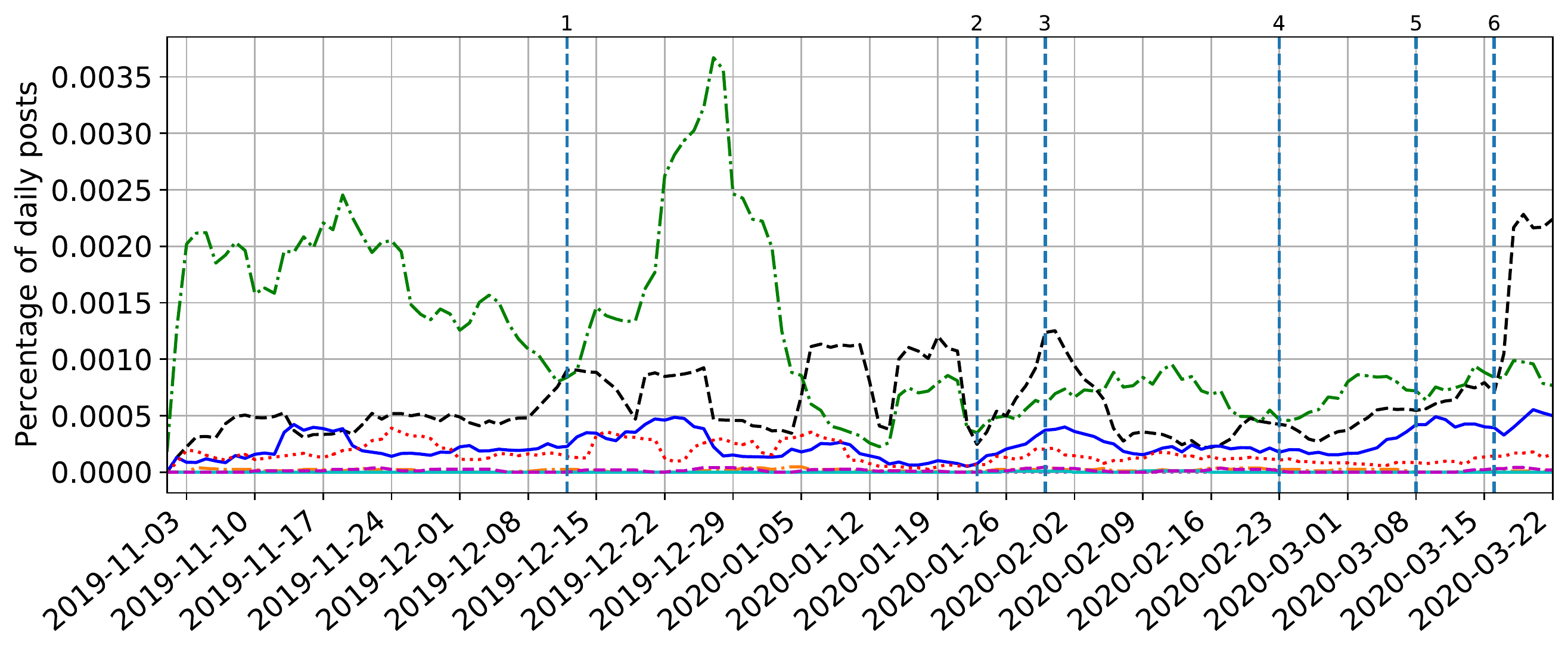}\label{fig:twitter_counts_other-slurs_rel}}
\caption{Mentions of Sinophobic racial slurs on Twitter. }
\label{fig:temporal_slurs_twitter}
\end{figure*}

\section{Temporal Analysis}

We start our analysis by studying the temporal dynamics of words related to ``china'' and ``chinese'' on 4chan's \dspol and Twitter.
Also, we investigate the prevalence of several racial slurs targeted towards Chinese people.

\subsection{Socio-Political Terms}

Figure~\ref{fig:temporal_china_4chan} shows the number of occurrences of ``china'' and ``chinese,'' and the proportion of posts containing these two words on 4chan's \dspol on a daily base.
We also annotate (with vertical lines) real-world events related to the COVID-19 pandemic (see Table~\ref{tab:events} for more details).

We first observe a sudden increase for both words around January 23, 2020, the day the Chinese government officially locked down the city of Wuhan marking the first large-scale effort in China to combat COVID-19.\footnote{NB: The COVID-19 name was not chosen by the World Health Organization (WHO) until a few weeks later on February 11, 2020~\cite{who_naming}.}
After the Wuhan lock-down, the popularity of ``china'' and ``chinese'' declines until the latter part of February, right around the time that COVID-19 cases started to appear en masse in Europe.

On 23 February (annotation 4 in the figure), 11 municipalities in Lombardy, Italy were put on lock-down in an attempt to slow the explosion of community spread cases, and we start to see the use of ``china'' and ``chinese'' rise again.
This rate increases dramatically around March 9th (annotation 5), which is when the Italian government extended the lock-down to the entirety of Italy.
The second peak comes around 16 March, when Donald Trump referred COVID-19 as ``Chinese Virus'' in a tweet.

On Twitter (see Figure~\ref{fig:temporal_china_twitter}) we see the same high level trend: discussion about ``china'' and ``chinese'' has a large up tick when Wuhan is locked down, and then declines until COVID-19 hits Europe.
There is one important difference however.
The amount of relative discussion on Twitter during the first peak is much lower than the level of discussion once Europe comes into play.

This may be due to the fact that discussion on Twitter is more geographically distributed, or that  4chan's \dspol is more easily inflamed by conspiracies and racism-related posts.
Social distance may work as one factor in illustrating the gap between two peaks. 
Referring to the perception of others ~\cite{trope2010construal, bogardus1933social, akerlof1997social, corrigan2001prejudice}, this perception can be elevated by a familiarity of cultural, nationality, ethics, education, occupation, etc. 
Geographically intimacy, as well as close cultural background, leads to higher attention on Covid- 19 outbreak in Europe than the lock-down in Wuhan. 
More analysis on this will be presented later in this paper.

\subsection{Racial Slurs}

Besides ``china'' and ``chinese,'' we also analyze the temporal dynamics of Sinophobic racial slurs on 4chan's \dspol and Twitter.
We pick a set of 8 Sinophobic slurs, including ``chink,'' ``bugland,'' ``chankoro,'' ``chinazi,'' ``gook,'' ``insectoid,'' ``bugmen,'' and ``chingchong.''
Some of them are well-known racial slurs towards Chinese and Asian people~\cite{sinophobia_wiki},  such as ``chink,'' ``chingchong,'' and ``gook.''
Others (e.g., ``bugland'') are based on preliminary results where we used word embeddings to discover other racial slurs (see Section~\ref{section:content_analysis} for more details).

The results are depicted in Figure~\ref{fig:temporal_slurs_4chan} and~\ref{fig:temporal_slurs_twitter} for \dspol and Twitter, respectively.
We again observe two peaks around 23 January and 16 March 2020 for these slurs on both platforms, similar to the results for ``china'' and ``chinese.''
In particular, at the highest points of these two peaks, more than 1\% of the total \dspol posts contain ``chink.''
This suggests that the COVID-19 crisis indeed drove the rise of Sinophobia on the Web.
Also, unlike what we observed on \dspol, on Twitter these Sinophobic slurs experienced a higher popularity around the middle of December, 2019.
An important factor for this might be that Donald Trump signed off a new trade deal between China and the USA during that time (see Table~\ref{tab:events}).

When comparing the popularity of these slurs across the two Web communities we find several differences.
On \dspol, we find that the most popular slur is ``chink,'' followed by ``gook'' and ``chingchong.'' 
On the other hand, for Twitter we observe that the most popular racial slur is ``chinazi'' which barely appears on \dspol. 
This highlights that there are fundamental differences across these platforms and that there is need to use more sophisticated techniques, rather than using a dictionary of terms, to capture the peculiarities and differences in the use of language across these two Web communities.

\begin{table*}[]
\centering
\resizebox{\textwidth}{!}{%
\begin{tabular}{@{}llllll|lrlrlr@{}}
\toprule
\multicolumn{6}{c|}{\textbf{4chan's /pol/}}                                                                                                                                                                                                                                                                                                                                                                                     & \multicolumn{6}{c}{\textbf{Twitter}}                                                                                                                                                                                                                                                                                                                                                                                                                                                        \\ \midrule
\textbf{\begin{tabular}[c]{@{}l@{}}Word\\ (China)\end{tabular}} & \textbf{\begin{tabular}[c]{@{}l@{}}Similarity\\ (China)\end{tabular}} & \textbf{\begin{tabular}[c]{@{}l@{}}Word\\ (Chinese)\end{tabular}} & \textbf{\begin{tabular}[c]{@{}l@{}}Similarity\\ (Chinese)\end{tabular}} & \textbf{\begin{tabular}[c]{@{}l@{}}Word\\ (Virus)\end{tabular}} & \textbf{\begin{tabular}[c]{@{}l@{}}Similarity\\ (Virus)\end{tabular}} & \textbf{\begin{tabular}[c]{@{}l@{}}Word\\ (China)\end{tabular}} & \multicolumn{1}{l}{\textbf{\begin{tabular}[c]{@{}l@{}}Similarity\\ (China)\end{tabular}}} & \textbf{\begin{tabular}[c]{@{}l@{}}Word\\ (Chinese)\end{tabular}} & \multicolumn{1}{l}{\textbf{\begin{tabular}[c]{@{}l@{}}Similarity\\ (Chinese)\end{tabular}}} & \textbf{\begin{tabular}[c]{@{}l@{}}Word\\ (Virus)\end{tabular}} & \multicolumn{1}{l}{\textbf{\begin{tabular}[c]{@{}l@{}}Similarity\\ (Virus)\end{tabular}}} \\ \midrule
chinas                                                          & 0.773                                                                 & chines                                                            & 0.830                                                                   & coronovirus                                                     & 0.846                                                                 & \textbf{chinese}                                                & 0.674                                                                                     & \textbf{taiwanese}                                                & 0.756                                                                                       & papilloma                                                       & 0.702                                                                                     \\
chinkland                                                       & 0.761                                                                 & chink                                                             & 0.818                                                                   & covid                                                           & 0.839                                                                 & destabl                                                         & 0.669                                                                                     & japane                                                            & 0.705                                                                                       & spr                                                             & 0.700                                                                                     \\
\textbf{chinese}                                                & 0.757                                                                 & chineese                                                          & 0.791                                                                   & coronavirus                                                     & 0.808                                                                 & \textbf{ccp}                                                    & 0.662                                                                                     & mainla                                                            & 0.677                                                                                       & viruse                                                          & 0.692                                                                                     \\
\textbf{ccp}                                                    & 0.748                                                                 & \textbf{china}                                                    & 0.757                                                                   & \textbf{corona}                                                 & 0.798                                                                 & uyghur                                                          & 0.661                                                                                     & turkistani                                                        & 0.676                                                                                       & mutate                                                          & 0.681                                                                                     \\
nk                                                              & 0.743                                                                 & \textbf{taiwanese}                                                & 0.752                                                                   & virius                                                          & 0.783                                                                 & fipa                                                            & 0.660                                                                                     & \textbf{china}                                                    & 0.674                                                                                       & \textbf{corona}                                                 & 0.678                                                                                     \\
chyna                                                           & 0.742                                                                 & wuhan                                                             & 0.725                                                                   & vrius                                                           & 0.782                                                                 & renminbi                                                        & 0.654                                                                                     & breifli                                                           & 0.671                                                                                       & transmissible                                                   & 0.670                                                                                     \\
bioattack                                                       & 0.736                                                                 & chinks                                                            & 0.719                                                                   & cornovirus                                                      & 0.781                                                                 & boycottbeij                                                     & 0.653                                                                                     & learnchinese                                                      & 0.671                                                                                       & ebol                                                            & 0.661                                                                                     \\
biowepon                                                        & 0.719                                                                 & ccp                                                               & 0.713                                                                   & wuflu                                                           & 0.780                                                                 & fentanylchina                                                   & 0.653                                                                                     & stillnoinfo                                                       & 0.668                                                                                       & mononucleosi                                                    & 0.657                                                                                     \\
chicomms                                                        & 0.718                                                                 & chinease                                                          & 0.701                                                                   & cornavirus                                                      & 0.779                                                                 & eastturkistan                                                   & 0.652                                                                                     & vietnamese                                                        & 0.667                                                                                       & desease                                                         & 0.655                                                                                     \\
bugland                                                         & 0.712                                                                 & chinamen                                                          & 0.699                                                                   & convid                                                          & 0.778                                                                 & xinj                                                            & 0.651                                                                                     & xijingp                                                           & 0.661                                                                                       & flue                                                            & 0.652                                                                                     \\
wuhan                                                           & 0.711                                                                 & japanese                                                          & 0.693                                                                   & hivs                                                            & 0.777                                                                 & xijinp                                                          & 0.647                                                                                     & manchuria                                                         & 0.660                                                                                       & wuhanflu                                                        & 0.651                                                                                     \\
chinkistan                                                      & 0.711                                                                 & korean                                                            & 0.690                                                                   & pathogen                                                        & 0.774                                                                 & falung                                                          & 0.647                                                                                     & putonghua                                                         & 0.660                                                                                       & coronar                                                         & 0.649                                                                                     \\
choyna                                                          & 0.696                                                                 & chingchong                                                        & 0.687                                                                   & supervirus                                                      & 0.773                                                                 & governemnt                                                      & 0.646                                                                                     & cambodian                                                         & 0.660                                                                                       & nucleotid                                                       & 0.648                                                                                     \\
chine                                                           & 0.682                                                                 & mainlander                                                        & 0.685                                                                   & disease                                                         & 0.764                                                                 & xijingp                                                         & 0.645                                                                                     & hainan                                                            & 0.658                                                                                       & pesti                                                           & 0.646                                                                                     \\
chiniggers                                                      & 0.682                                                                 & chinaman                                                          & 0.680                                                                   & sars                                                            & 0.760                                                                 & chinazi                                                         & 0.644                                                                                     & pribumi                                                           & 0.655                                                                                       & chikungunya                                                     & 0.646                                                                                     \\
tradewar                                                        & 0.682                                                                 & cpc                                                               & 0.678                                                                   & viruse                                                          & 0.759                                                                 & xinjiang                                                        & 0.644                                                                                     & kazakh                                                            & 0.653                                                                                       & conoravirus                                                     & 0.645                                                                                     \\
nambawan                                                        & 0.681                                                                 & chicom                                                            & 0.676                                                                   & biowepon                                                        & 0.757                                                                 & ccpchina                                                        & 0.639                                                                                     & prc                                                               & 0.651                                                                                       & commens                                                         & 0.644                                                                                     \\
koreas                                                          & 0.680                                                                 & mainlanders                                                       & 0.674                                                                   & asymptomic                                                      & 0.754                                                                 & jinpin                                                          & 0.637                                                                                     & qingpu                                                            & 0.650                                                                                       & protozoa                                                        & 0.644                                                                                     \\
chynah                                                          & 0.678                                                                 & chinkland                                                         & 0.673                                                                   & sras                                                            & 0.753                                                                 & beltandroad                                                     & 0.630                                                                                     & laotian                                                           & 0.644                                                                                       & dengue                                                          & 0.642                                                                                     \\
chines                                                          & 0.675                                                                 & shina                                                             & 0.671                                                                   & megavirus                                                       & 0.750                                                                 & eastturkestan                                                   & 0.630                                                                                     & shandong                                                          & 0.643                                                                                       & antibi                                                          & 0.642                                                                                     \\ \bottomrule
\end{tabular}%
}
\caption{Top 20 most similar words to the words ``china,'' ``chinese,'' and ``virus'' obtained from the word2vec models trained for the whole period (November 2019 - March 2020).}
\label{tab:top_20_whole}
\end{table*}

\subsection{Discussion}

A common theme among racist ideology is that of an invading virus.
History is rife with examples of diseases being attributed to specific races and nationalities, and there is no reason to believe that COVID-19 would buck this trend; the first identified COVID-19 cases \emph{did} originate in China.
However, the world today is much more diverse and connected than it was in the 15th century when Italians dubbed syphilis the ``French disease.''

Figures~\ref{fig:temporal_china_4chan} and~\ref{fig:temporal_china_twitter} make it quite clear that 4chan and Twitter are heavily discussing China in relation to COVID-19, and that this discussion accelerated rapidly once the Western world became affected.
The upswing is potentially related to the scapegoating phenomenon~\cite{toker1972scapegoat}
The first cases originated in China, and the Chinese government was the first to take active and serious measures to combat its spread prompting a reasonable degree of discussion.
When these measures were ineffective in preventing the spread to the Western world, however, China's existing association with COVID-19, in particular China's ``failure'' to prevent its spread make it a \emph{just} scapegoat~\cite{allport1954nature} in the face of a looming pandemic.

That said, we do see meaningful differences in the use of \emph{slurs} on \dspol and Twitter.
\dspol's use of slurs tracks with the use of ``china'' and ``chinese'' to a worrying degree, but this is much less pronounced on Twitter.
This is not entirely unsurprising considering that \dspol is well known to be a locus of racist ideology, however it is worthwhile discussing some of the theory around \emph{why} it tracks so well.
The clearly racist reaction fits the notion of \emph{defensive denial}, which is a common strategy for coping with stress~\cite{anderson2013diagnosing,dohrenwend2004positive,houston1973viability,janoff1987coping,sirin2015discrimination}.
Essentially, the early stages of COVID-19 were exclusively a \emph{Chinese} problem; ``superior'' Western society had nothing to worry about, even though experts were warning of a pandemic breakout even before Wuhan was locked down.
This conforms with the \emph{scapegoating} theory of clinical psychology, in which members of a group project unwanted self aspects onto another person or group, then attack the scapegoat believing that ``this is not me''~\cite{clark2002scapegoating,gemmill1989dynamics,prudhomme1970reflections}.
Political scientists have argued that scapegoating is a major driver for racism in a number of settings~\cite{denike2015scapegoat,patton2007any}.

\section{Content Analysis}
\label{section:content_analysis}

\subsection{Method}

To analyze the content, more specifically the \emph{context} of the use of specific words, we train multiple word2vec models~\cite{mikolov2013distributed} for each Web community.
In a nutshell, these models map words into a high-dimensional vector space so that words that are used in a similar way are close to each other.
To do this, we leverage the skip-gram model, which is a shallow neural network aiming to predict the context of a specific word.
In this work, we train three groups of word2vec models for each of Twitter and \dspol:
\begin{compactenum}
\item One word2vec model (\wvall{}) trained on all posts made during the period between 28 October, 2019 and 22 March, 2020.
We denote the period by \duration{}.
This model allows us to study the use of words for the entire duration of our study.
\item One distinct word2vec model for each week between 28 October, 2019 and 22 March, 2020, denoted by \wvweekly{} 
($i$ is the $i$th week in \duration{}).
These models allow us to study \emph{changes} in the use of words over time.
\item One word2vec model trained on historical data for all posts shared between July 1, 2016 and November 1, 2019 (\wvcontrol{}).
This model acts as a baseline and allows us to investigate the emergence of new terms during the period of our study.
\end{compactenum}

\begin{figure*}[t!]
\centering
\includegraphics[width=0.8\textwidth]{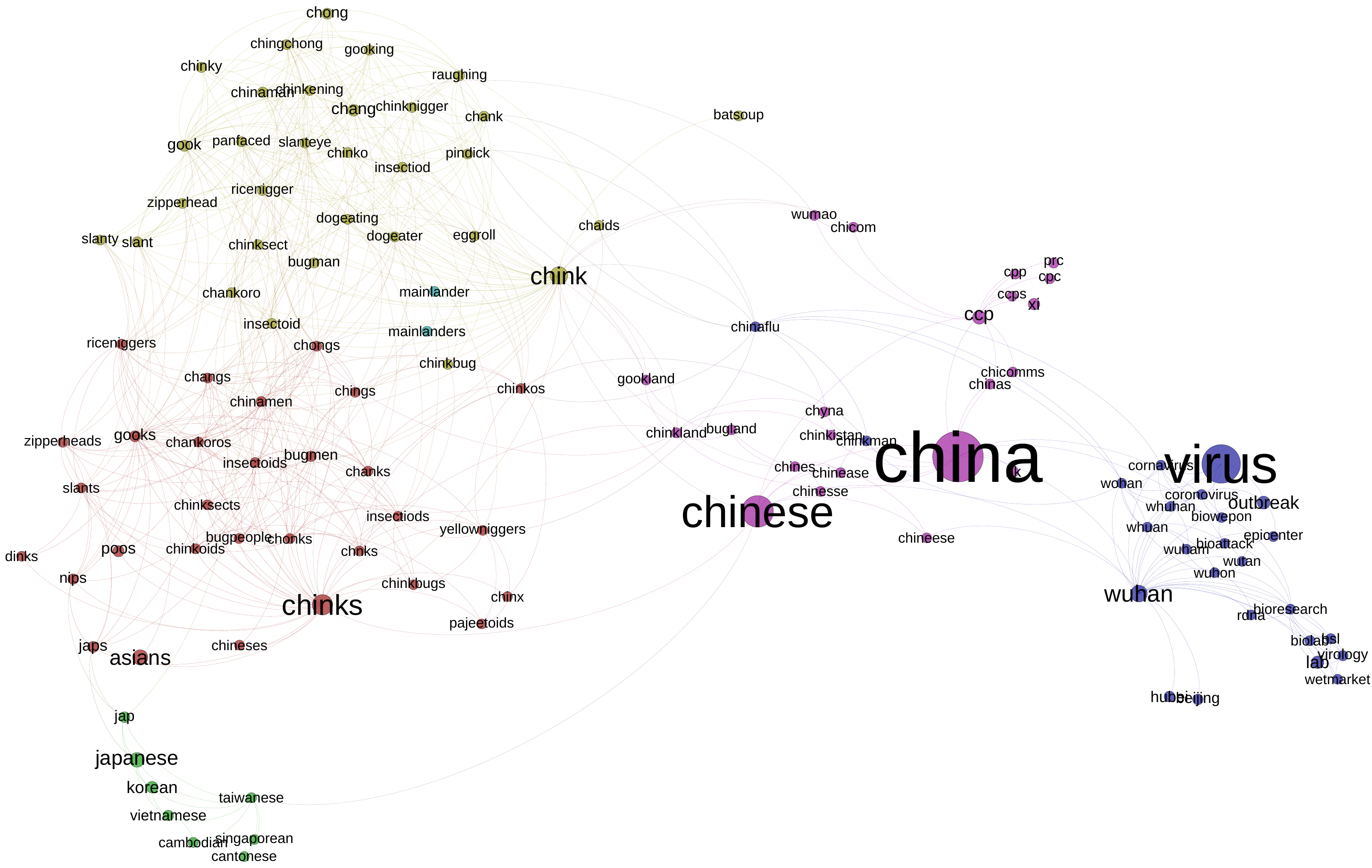}
\caption{Visualization of a 2-hop graph from the word ``chinese'' on 4chan's \dspol.}
\label{fig:graph_chinese_4chan}
\end{figure*}

\subsection{Exploring the Context of Terms}

First, we look into the overall use of words on 4chan's \dspol using the word2vec model trained on the period between 28 October, 2019 and 22 March, 2020 (\wvall{}).
In this model, words used in similar context will present similar vectors.
The left side of Table~\ref{tab:top_20_whole} reports the top 20 most similar words for the terms ``china,'' ``chinese,'' and ``virus.''
We make several observations:
first, we note that there are many derogatory terms for Asian people, Chinese people in particular, in the top 20 most similar terms.
Some examples include ``chink'' (derogatory term referring to Asian people), ``chinkland'' (referring to the land of chinks, i.e., China), and ``chiniggers'' (an offensive word created by combining ``china'' and ``nigger'').
For instance, a \dspol user posted: \emph{``We should have never let these Chiniggers into the country or enforced a mandatory quarantine for anyone coming from contaminated areas. But it's too late now.''}
Another \dspol user posted: \emph{``You chinks deserve it, there's no shithole of a country that could be as disgusting as chinkland.''}
This indicates that \dspol users use a wide variety of derogatory terms to possibly disseminate hateful ideology towards Chinese and Asian people.
Second, by looking at the most similar words of the term ``virus,'' we find several terms related to the COVID-19 pandemic~\cite{coronavirus_pandemic}.
This is evident since the four most similar words to the term ``virus'' are related to COVID-19, specifically, ``coronovirus,'' ``covid,'' ``coronavirus,'' and ``corona.''
This indicates that the overall use of words in \dspol is highly affected by the COVID-19 pandemic, and this event is likely to cause changes in the use of language by users.

The corresponding results for Twitter is shown on the right side of Table~\ref{tab:top_20_whole}.
On Twitter we observe multiple political-related terms that are similar to ``china'' and ``chinese,'' such as ``government'' and ``ccp'' (Chinese Communist Party).
Furthermore, we again observe, some potentially offensive terms like ``chinazi,'' which indicates that the use of Sinophobic content is not limited to fringe Web communities like 4chan, and it also exists in mainstream Web communities like Twitter.
Also, many terms that are similar to ``virus'' are also related to COVID-19, such as ``corona'' and ``coronavirus.''
This indicates Twitters users' word usage are influenced by the COVID-19 pandemic as well.

\begin{figure*}[t!]
\centering
\includegraphics[width=0.8\textwidth]{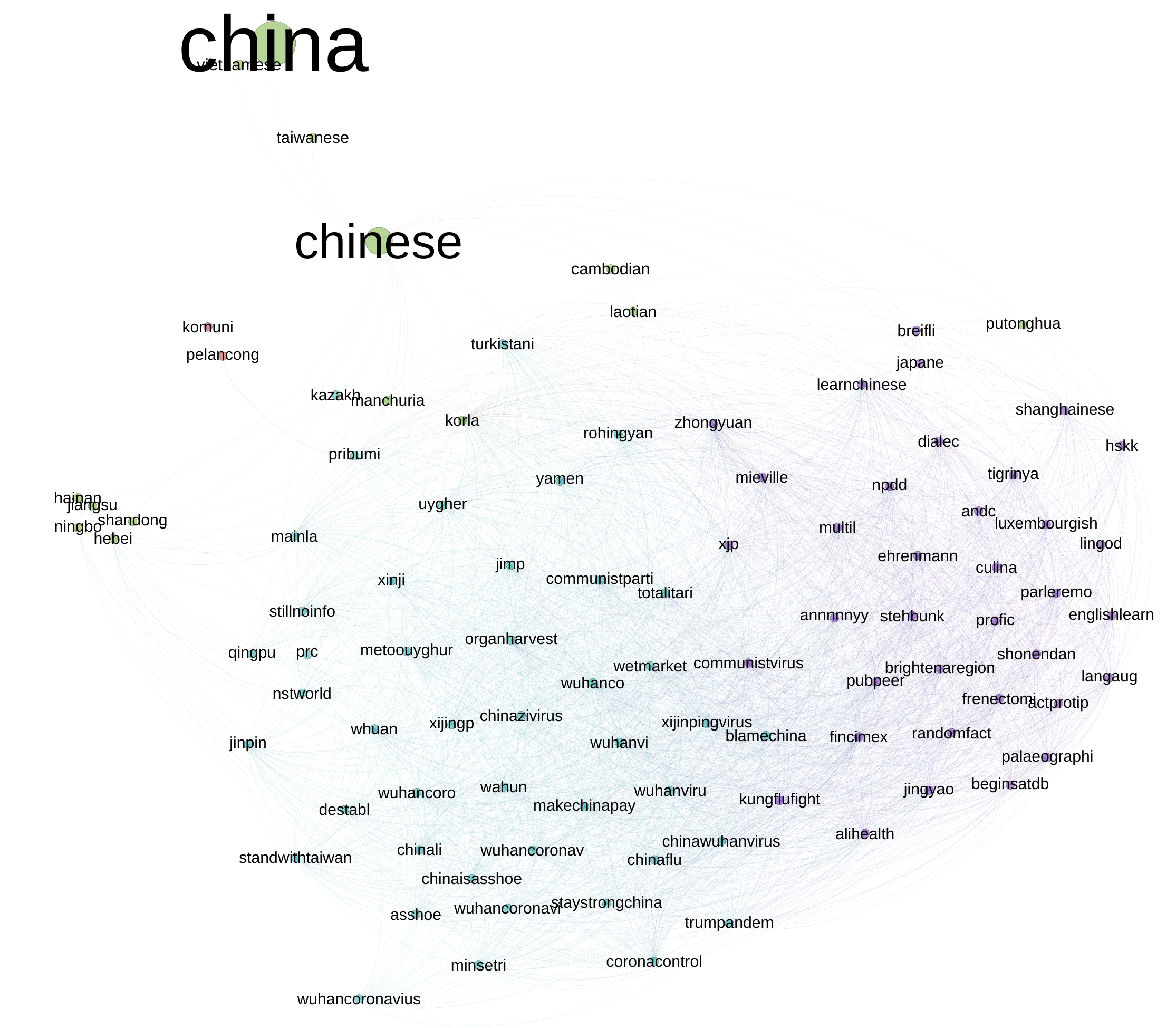}
\caption{Visualization of a 2-hop graph from the word ``chinese'' on Twitter.}
\label{fig:graph_chinese_twitter}
\end{figure*}

\subsection{Visualizing the Similarity between the Use of Terms}

To better visualize the use of language related to Chinese people, we create graphs that visualize the use of words that are similar to the term ``chinese,'' following the methodology by Zannettou et al.~\cite{zannettou2020quantitative}.
In a nutshell, we create a graph where nodes are words and an edge between the words exists if their cosine similarity (obtained from the trained word2vec model) is above a pre-defined threshold.\footnote{The threshold differs for each resulted graph in a way that it maximizes the readability of the graph.} 
We limit the graph into nodes that are two hops away from a specific word of interest (in this case ``chinese'').
Then, we perform various tasks for visualizing the graph.
First, the graph is layed out in the graph space with an algorithm that takes into account the weights of the edges~\cite{jacomy2014forceatlas2}.
That is, words that have large cosine similarities are layed out closer in the graph space.
Second, the size of each node is relative to the number of occurrences of the word in our dataset.
Third, we run the the Louvain community detection method~\cite{blondel2008fast} on the graph and represent nodes that belong to the same community with the same color. 
The resulting graphs are depicted in Fig~\ref{fig:graph_chinese_4chan} and Fig~\ref{fig:graph_chinese_twitter} for \dspol and Twitter, respectively.

By inspecting the obtained communities of words in Figure~\ref{fig:graph_chinese_4chan}, we observe several interesting themes around the use of words related to ``chinese.''
First, we observe a community that is highly related to the COVID-19 pandemic (blue community on bottom right). 
Interestingly, within this community, we also observe terms like ``biowepon'' (sic) and ``bioattack,'' likely indicating that \dspol users are sharing probably false information about the pandemic, for instance claiming that the whole pandemic is a ``bioattack'' from the Chinese on the Western world.
For example, a \dspol user posted: \emph{``Anyone that doesn't realize this is a Chinese bioweapon by now is either a brainlet or a chicom noodle nigger.''}
Second, we observe two tightly-knit communities (red and yellow communities on left-side of the graph) that appear to predominantly include derogatory terms towards Asian, and in particular Chinese people.
Some of the words in these communities are ``ricenigger,'' ``chinksect,'' ``chankoro,'' ``chinks,'' ``yellowniggers,'' and ``pindick.''
By looking at some examples of posts from \dspol users, we observe the use of these terms for disseminating hate: e.g., \emph{``Chang you useless ricenigger fuck off. Just call the bitch and ask her youll see this is fucking ccp bs. ITS A FUCKING EXPERIMENTAL CHINK BIOWEP''} and \emph{``I fucking hate chinks. Stop spreading viruses everywhere you pindick cunts.''}
Interestingly, the most distant word in these communities is the word ``batsoup,'' which is closer to the community related to COVID-19~\cite{batsoup_misinformation}.
The rest of the communities in this graph are seemingly related to China in general (purple community) and to other countries in Asia (green community).
Overall, this graph highlights that \dspol users use a wide variety of derogatory terms to characterize Chinese people.

When looking at the graph obtained for Twitter (see Figure~\ref{fig:graph_chinese_twitter}), we observe an interesting community of terms (blue), which includes words related to the COVID-19 pandemic. 
We observe a large number of words that are seemingly anti-China like the terms ``makechinapay,'' ``blamechina,'' and ``chinaisasshoe.''
At the same time, there are a lot of terms referring to the virus itself like ``chinawuhanvirus,'' ``chinaflu,'' and ``coronacontrol,'' as well as a few terms that aim to support Chinese people through this crisis like ``staystrongchina.''
For example, a Twitter user posted: \emph{``How do you say “Chi-com asshoe”? \#ChinesePropaganda \#ChinaLiedPeopleDied \#ChinaVirus \#WuhanCoronavirus.''}
The other communities on the graph include various terms related to happenings in China and other Asian countries/regions.

\subsection{Discussion}

By taking a deeper look at profanities that appeared among the terms, we can roughly divide them into two groups: one is insults addressing Asian people, such as racist variations of ``china'' and ``chinese'' (e.g., ``chinkland,'' ``chingchong,'' and ``chinksect'') or culturally oriented racist terms, including attacking dietary habits (e.g., ``ricenigger''), skin tone (e.g., ``yellownigger''), or sexual stereotypes (e.g., ``pindick'').
The frequent appearance of swear words among the terms can indicate an abreaction to the rising fear and stress in front of the disease~\cite{guvendir2015males,dewaele2004emotional}.
At the same time, the racist and targeted focus of these slurs can be explained with the mechanism of \emph{defensive aggression}, either focusing on cultural taboos, such as sexuality~\cite{dewaele2004emotional,hughes1998swearing}, or perpetuating societal oppression~\cite{hughes1998swearing}.

\begin{table*}[]
\centering
\resizebox{\textwidth}{!}{%
\begin{tabular}{@{}lrlrlr|lrlrlr@{}}
\toprule
\multicolumn{6}{c|}{\textbf{First Word2vec model (week ending on 2019/11/03)}}                                                                                                                                                                                                                                                                                                                                                                                                               & \multicolumn{6}{c}{\textbf{Last Word2vec model (week ending on 2020/03/22)}}                                                                                                                                                                                                                                                                                                                                                                                                                \\ \midrule
\textbf{\begin{tabular}[c]{@{}l@{}}Word\\ (china)\end{tabular}} & \multicolumn{1}{l}{\textbf{\begin{tabular}[c]{@{}l@{}}Similarity\\ (china)\end{tabular}}} & \textbf{\begin{tabular}[c]{@{}l@{}}Word\\ (chinese)\end{tabular}} & \multicolumn{1}{l}{\textbf{\begin{tabular}[c]{@{}l@{}}Similarity\\ (chinese)\end{tabular}}} & \textbf{\begin{tabular}[c]{@{}l@{}}Word\\ (virus)\end{tabular}} & \multicolumn{1}{l|}{\textbf{\begin{tabular}[c]{@{}l@{}}Similarity\\ (virus)\end{tabular}}} & \textbf{\begin{tabular}[c]{@{}l@{}}Word\\ (china)\end{tabular}} & \multicolumn{1}{l}{\textbf{\begin{tabular}[c]{@{}l@{}}Similarity\\ (china)\end{tabular}}} & \textbf{\begin{tabular}[c]{@{}l@{}}Word\\ (chinese)\end{tabular}} & \multicolumn{1}{l}{\textbf{\begin{tabular}[c]{@{}l@{}}Similarity\\ (chinese)\end{tabular}}} & \textbf{\begin{tabular}[c]{@{}l@{}}Word\\ (virus)\end{tabular}} & \multicolumn{1}{l}{\textbf{\begin{tabular}[c]{@{}l@{}}Similarity\\ (virus)\end{tabular}}} \\ \midrule
japan                                                           & 0.779                                                                                     & han                                                               & 0.740                                                                                       & infectious                                                      & 0.700                                                                                      & ccp                                                             & 0.738                                                                                     & ccp                                                               & 0.744                                                                                       & bioengineered                                                   & 0.726                                                                                     \\
singapore                                                       & 0.737                                                                                     & china                                                             & 0.733                                                                                       & viruses                                                         & 0.682                                                                                      & chinese                                                         & 0.730                                                                                     & chink                                                             & 0.730                                                                                       & wuflu                                                           & 0.722                                                                                     \\
chinese                                                         & 0.733                                                                                     & tibetans                                                          & 0.656                                                                                       & infects                                                         & 0.680                                                                                      & chinas                                                          & 0.711                                                                                     & china                                                             & 0.730                                                                                       & decease                                                         & 0.714                                                                                     \\
ccp                                                             & 0.730                                                                                     & chinks                                                            & 0.641                                                                                       & inject                                                          & 0.665                                                                                      & chinks                                                          & 0.694                                                                                     & chinks                                                            & 0.662                                                                                       & covid                                                           & 0.712                                                                                     \\
taiwan                                                          & 0.711                                                                                     & japan                                                             & 0.633                                                                                       & pathogen                                                        & 0.652                                                                                      & chink                                                           & 0.662                                                                                     & ebright                                                           & 0.635                                                                                       & supervirus                                                      & 0.698                                                                                     \\
russia                                                          & 0.708                                                                                     & singapore                                                         & 0.630                                                                                       & ebola                                                           & 0.650                                                                                      & chinkland                                                       & 0.651                                                                                     & transparently                                                     & 0.631                                                                                       & viruses                                                         & 0.695                                                                                     \\
india                                                           & 0.703                                                                                     & taiwanese                                                         & 0.627                                                                                       & lice                                                            & 0.630                                                                                      & whistleblowers                                                  & 0.641                                                                                     & taiwanese                                                         & 0.631                                                                                       & specimens                                                       & 0.689                                                                                     \\
venezuela                                                       & 0.697                                                                                     & chink                                                             & 0.622                                                                                       & vectors                                                         & 0.623                                                                                      & chernobyl                                                       & 0.635                                                                                     & amerikkkans                                                       & 0.627                                                                                       & mutated                                                         & 0.684                                                                                     \\
surpass                                                         & 0.691                                                                                     & filipinos                                                         & 0.618                                                                                       & spreads                                                         & 0.618                                                                                      & childkiller                                                     & 0.634                                                                                     & chinkoid                                                          & 0.622                                                                                       & corvid                                                          & 0.681                                                                                     \\
korea                                                           & 0.670                                                                                     & payback                                                           & 0.615                                                                                       & malignant                                                       & 0.614                                                                                      & embargo                                                         & 0.626                                                                                     & mainlanders                                                       & 0.620                                                                                       & virulence                                                       & 0.675                                                                                     \\
opium                                                           & 0.660                                                                                     & ccp                                                               & 0.613                                                                                       & outbreak                                                        & 0.613                                                                                      & chyna                                                           & 0.626                                                                                     & zainichi                                                          & 0.619                                                                                       & inoculated                                                      & 0.674                                                                                     \\
mainland                                                        & 0.654                                                                                     & mainland                                                          & 0.611                                                                                       & disposed                                                        & 0.611                                                                                      & nk                                                              & 0.624                                                                                     & bioterrorism                                                      & 0.618                                                                                       & corona                                                          & 0.674                                                                                     \\
geostrategic                                                    & 0.654                                                                                     & cantonese                                                         & 0.611                                                                                       & deficiency                                                      & 0.611                                                                                      & chankoro                                                        & 0.623                                                                                     & labelling                                                         & 0.613                                                                                       & disease                                                         & 0.673                                                                                     \\
surpassed                                                       & 0.653                                                                                     & koreans                                                           & 0.611                                                                                       & ensues                                                          & 0.607                                                                                      & sanctions                                                       & 0.619                                                                                     & cccp                                                              & 0.612                                                                                       & chimera                                                         & 0.671                                                                                     \\
steamrolled                                                     & 0.650                                                                                     & hui                                                               & 0.608                                                                                       & drawings                                                        & 0.604                                                                                      & numba                                                           & 0.619                                                                                     & originates                                                        & 0.611                                                                                       & flu                                                             & 0.670                                                                                     \\
hk                                                              & 0.649                                                                                     & manchuria                                                         & 0.607                                                                                       & carnage                                                         & 0.603                                                                                      & asshoe                                                          & 0.616                                                                                     & ideia                                                             & 0.608                                                                                       & transmittable                                                   & 0.669                                                                                     \\
indonesia                                                       & 0.648                                                                                     & paramount                                                         & 0.604                                                                                       & disposition                                                     & 0.603                                                                                      & retaliate                                                       & 0.613                                                                                     & spies                                                             & 0.606                                                                                       & infection                                                       & 0.668                                                                                     \\
pakistan                                                        & 0.647                                                                                     & mandarin                                                          & 0.603                                                                                       & bioengineering                                                  & 0.602                                                                                      & foothold                                                        & 0.612                                                                                     & chines                                                            & 0.605                                                                                       & transmits                                                       & 0.667                                                                                     \\
manchuria                                                       & 0.645                                                                                     & japs                                                              & 0.602                                                                                       & abduction                                                       & 0.596                                                                                      & velllly                                                         & 0.609                                                                                     & wumao                                                             & 0.605                                                                                       & chickenpox                                                      & 0.666                                                                                     \\
chinks                                                          & 0.644                                                                                     & dravidians                                                        & 0.598                                                                                       & contagion                                                       & 0.596                                                                                      & engrish                                                         & 0.606                                                                                     & westerner                                                         & 0.604                                                                                       & reinfecting                                                     & 0.665                                                                                     \\ \bottomrule
\end{tabular}%
}
\caption{Top 20 most similar words to the words ``china,'' ``chinese,'' and ``virus'' for the first and last trained word2vec models from 4chan's \dspol.}
\label{tab:top_20_first_last}
\end{table*}

\begin{figure*}[t!]
\center
\subfigure[$\mathcal{W}_{t=0}$]{\includegraphics[width=0.8\textwidth]{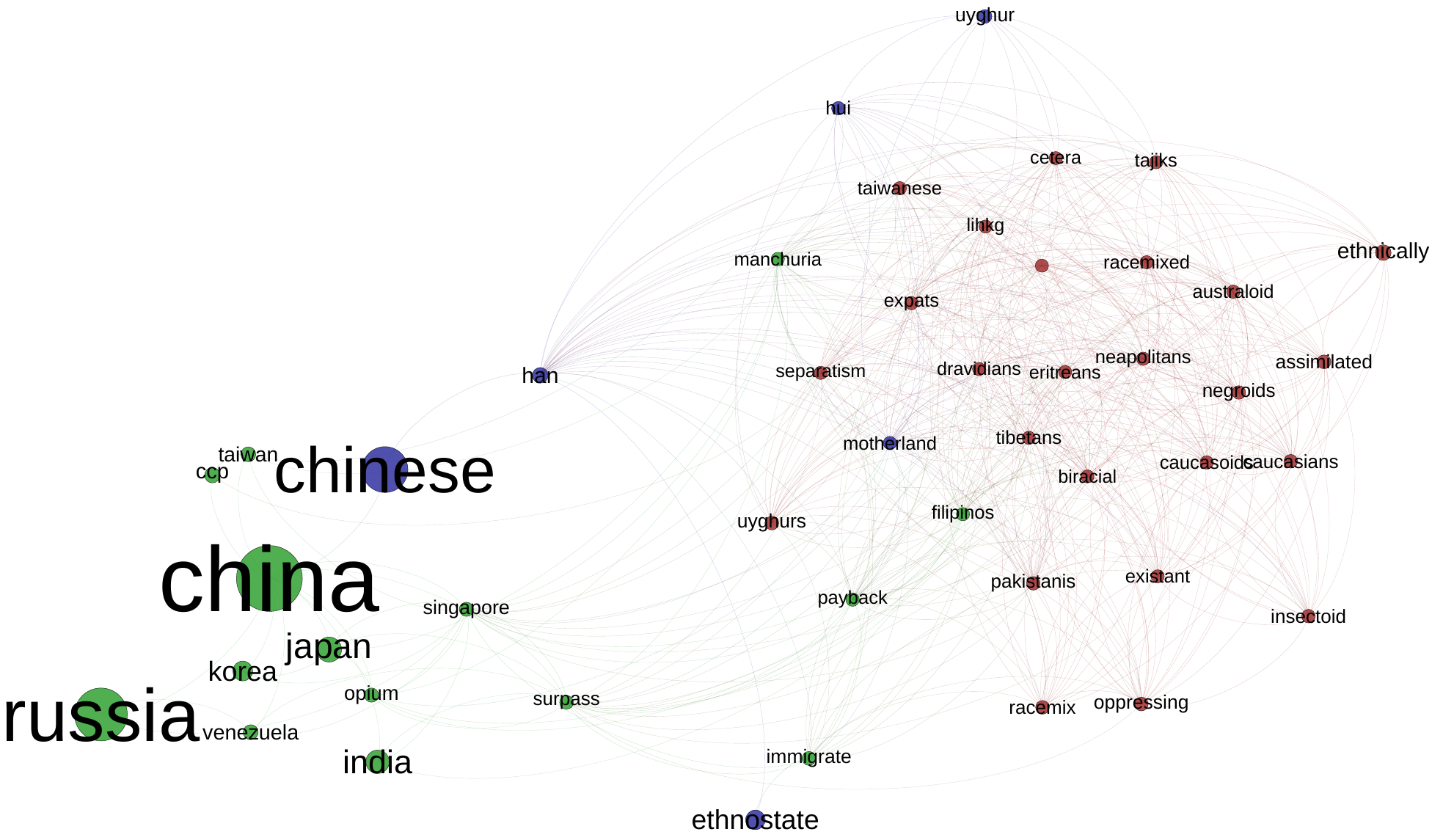}\label{fig:graph_chinese_4chan_first}}
\subfigure[$\mathcal{W}_{t=-1}$]{\includegraphics[width=0.8\textwidth]{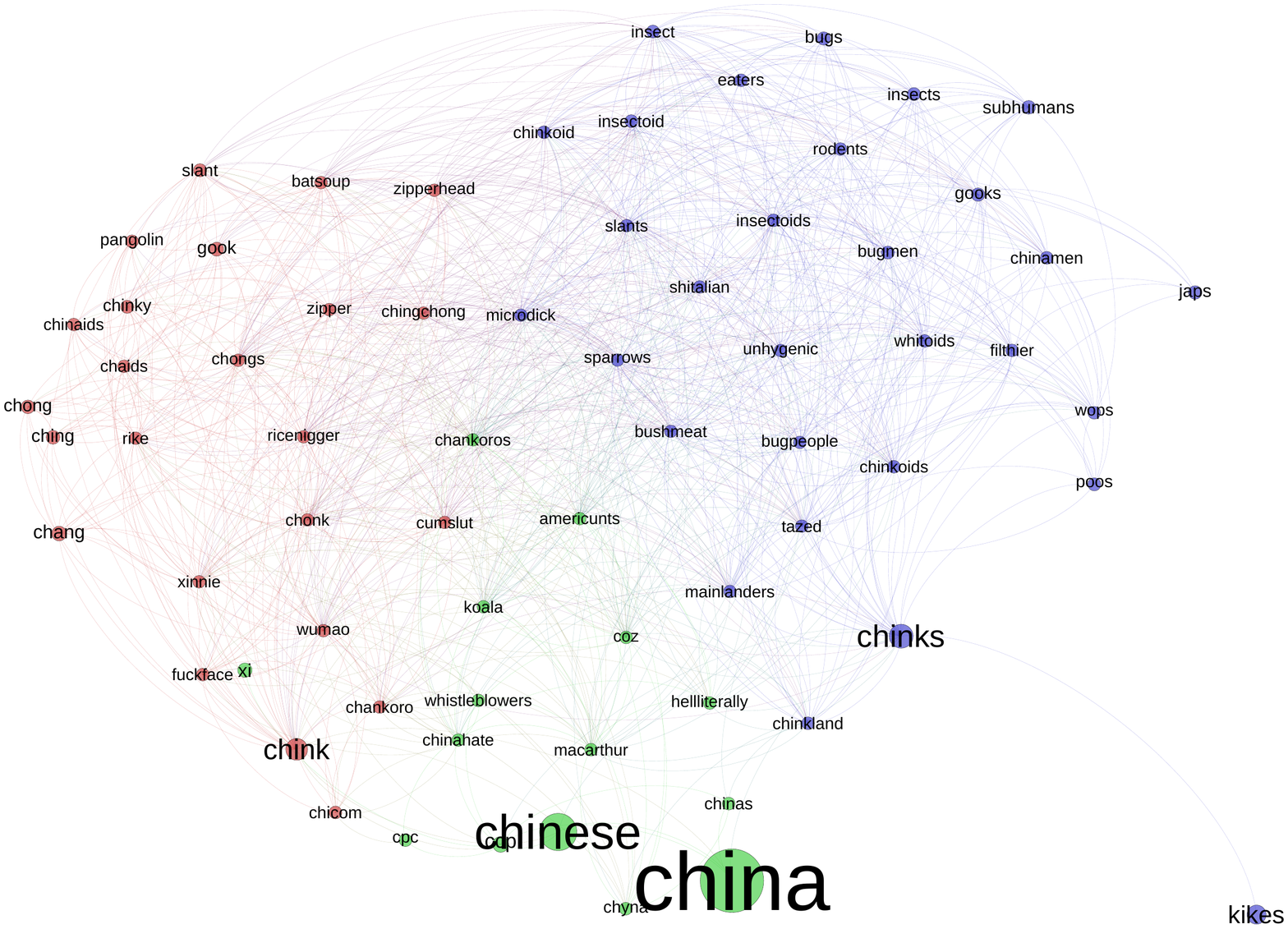}\label{fig:graph_chinese_4chan_last}}
\caption{Visualization of a 2-hop graph from the word ``chinese'' on 4chan's \dspol using the first and last weekly word2vec models.} 
\label{fig:graph_chinese_4chan_overtime}
\end{figure*}

\section{Content Evolution}

Discussions on Web communities like 4chan's \dspol and Twitter are highly dynamic and respond to real-world events as they unfold.
Thus, we expect users on these Web communities to discuss various topics related to the COVID-19 pandemic.
Moreover, events like the COVID-19 pandemic unfold over time, and this is reflected by the dynamics of discussion on Web communities.

In previous sections we explored both the usage of some key terms related to Sinophobia, as well as a static understanding of content.
However, these previous analyses do not help us understand how Sinophobic language is evolving over time.
More specifically, there is a lack of understanding on how the context in which words are used changes, and also how new words are created.
The former is important because it provides significant insights into the scope and breadth of the problem.
The latter is important because the language of online extremism has been shown to include memes and slang that have completely contradictory meanings to ``normal'' usage, or do not even exist outside of the communities that use them.
We first study the Sinophobic language evolution on 4chan's \dspol, and in Section~\ref{section:evolve_twitter}, we will focus on Twitter.

\subsection{Evolution on 4chan's \dspol}
\label{section:evolve_4chan}

To study the evolution of discussions and use of language, we make use of the weekly word2vec models (\wvweekly{}).
To illustrate how these models are helpful, we initially compare the results from the model trained on the first week of our dataset ($\mathcal{W}_{t=0}$) with the model trained on the last week of our dataset ($\mathcal{W}_{t=-1}$).
Table~\ref{tab:top_20_first_last} reports the top 20 similar words to ``china,'' ``chinese,'' and ``virus,'' for the first and last weekly word2vec models (similar to how Table~\ref{tab:top_20_whole} shows results for a model trained on the entirety of our dataset).
Interestingly, we observe major differences between the most similar words obtained from the first and last models (comparing left sides of the Table with the right side), as well as between the whole model and these two weekly models (cf. Table~\ref{tab:top_20_whole} and Table~\ref{tab:top_20_first_last}).

We make several key observations.
First, when looking at the most similar words to the term ``china'' from the first week model (left side of Table~\ref{tab:top_20_first_last}), we observe words referring to other counties, mostly in Asia (e.g., ``japan,'' ``singapore,'' etc.), but also that the derogatory term ``chinks'' is among the top 20.
This result indicates that 4chan's \dspol users typically use racial slurs targeted to Chinese people, and this was also happening even before the outbreak of the COVID-19 pandemic.
Similar findings can be observed by looking at the most similar words to the term ``chinese.'' 
We observe the existence of racial slurs like ``chink,'' however, most of the other words relate to people originating from other Asian countries, such as ``koreans.''
When looking at the most similar words to the term ``virus,'' before the COVID-19 pandemic, we observe general terms related to diseases or other outbreaks, e.g., ``ebola.''

Second, by comparing the most similar words from the first and last models, we observe several interesting differences.
By looking at the most similar words to the term ``china,'' we observe that derogatory terms like ``chink'' have a higher cosine similarity compared to the first model, likely indicating a rise in the use of this term in discussions related to China.
Furthermore, we observe terms like ``chernobyl,'' which may indicate that \dspol users are comparing this outbreak with the Chernobyl disaster.
For example, a \dspol user posted: \emph{``I can see China collapsing after all this, just as the Chernobyl incident was the beginning of the end for the USSR....''}
We also see the term ``childkiller,'' which upon manual investigation is due to a particularly active user repeatedly posting that China created COVID-19 as a bioweapon.
Specifically, we find multiple occurrences of the following sentence in multiple \dspol posts: \emph{``CHINA CREATED THE CHINA BIOWEAPON MURDER DEATH CHILDKILLER VIRUS IN CHINA!''}
Interestingly, we also find some terms that seem to be sarcastic towards the way that Chinese people talk English.
For instance, the term ``numba'' refers to the word ``number'' and ``asshoe'' refers to the term ``asshole.''
Some examples from \dspol posts are: \emph{``Don’t trust China, China is asshoe''} and \emph{``TAIWAN NUMBA 1 CHINA NUMBA NONE!''}

Third, by looking into the most similar words to the term ``chinese,'' we observe the term ``bioterrorism'' likely indicating that 4chan's \dspol users are calling Chinese people as bioterrorists that is likely related to conspiracy theories that COVID-19 was bioengineered.
For example, a \dspol user posted: \emph{``THIS IS BIOTERRORISM NUKE CHINA NOW.''}
By looking at the most similar words to the term ``virus,'' we find that the most similar one is the term ``bioengineered,'' indicating that the conspiracy theory went viral on \dspol during that specific week and was discussed extensively.
For instance a \dspol user posted: \emph{``The bat soup is just a cover-up. One of (((Leiber)))'s chinks stole the bioengineered virus \& tried to patent it in China, violating export-controlled laws \& committing espionage. My guess is, he didn't handle the virus correctly, got himself sick, then infected others in the Wuhan wet market.''}
Finally, by looking at the other similar words to the term ``virus,'' we clearly observe those that are related to the COVID-19 pandemic with terms like ``wuflu'' (created by combining Wuhan and Flu), ``covid,'' and ``corona.''
For instance, a \dspol user posted \emph{``Die to wuflu already, boomers.''}
 
These differences are also more evident by looking at the graph visualizations in Figure~\ref{fig:graph_chinese_4chan_overtime}. 
To create these graphs, we use the same methodology as Figure~\ref{fig:graph_chinese_4chan}, for the first and last weekly trained word2vec models, visualizing the two-hop neighborhood of the term ``chinese.''
Looking at the graph obtained from the first model (see Figure~\ref{fig:graph_chinese_4chan_first}), we observe mostly innocuous terms related to Chinese people and other Asian people. 
By looking at the graph obtained from the last model (see Figure~\ref{fig:graph_chinese_4chan_last}), however, we observe an entirely different, more hateful behavior.
Specifically, the two main tightly-knit communities (red and blue communities), are filled with slurs used against Chinese people like ``ricenigger,'' ``fuckface,'' ``zipperhead,'' ``bugpeople,'' ``subhumans,'' etc.
Example of posts from \dspol include: ``\emph{I hope you fucking die in hell, you psychopathic zipperhead. You and your whole disgusting race}'' and ``\emph{We should unironically nuke China. Kill some bugpeople and eradicate COVID-19 at the same time.}''

Overall, these findings indicating that we are experiencing an explosion in the use of Chinese derogatory terms in fringe Web communities like 4chan's \dspol, in particular after the outbreak of the COVID-19 pandemic. 
These findings are particularly worrisome, since it is likely that as the pandemic evolves, it is likely to have further rise in the dissemination of racist and hateful ideology towards Chinese people that might also have real-world consequences, such as physical violence against Chinese people.

\begin{table*}[]
\centering
\resizebox{\textwidth}{!}{%
\begin{tabular}{@{}lrlrlr|lrlrlr@{}}
\toprule
\multicolumn{6}{c|}{\textbf{First word2vec model (week ending on 2019/11/03)}}                                                                                                                                                                                                                                                                                                                                                                                                                                                                           & \multicolumn{6}{c}{\textbf{Last word2vec model (week ending on 2020/03/22)}}                                                                                                                                                                                                                                                                                                                                                                                                                                                                            \\ \midrule
\multicolumn{1}{c}{\textbf{\begin{tabular}[c]{@{}c@{}}Word\\ (china)\end{tabular}}} & \multicolumn{1}{c}{\textbf{\begin{tabular}[c]{@{}c@{}}Similarity\\ (china)\end{tabular}}} & \multicolumn{1}{c}{\textbf{\begin{tabular}[c]{@{}c@{}}Word\\ (chinese)\end{tabular}}} & \multicolumn{1}{c}{\textbf{\begin{tabular}[c]{@{}c@{}}Similarity\\ (chinese)\end{tabular}}} & \multicolumn{1}{c}{\textbf{\begin{tabular}[c]{@{}c@{}}Word\\ (virus)\end{tabular}}} & \multicolumn{1}{c|}{\textbf{\begin{tabular}[c]{@{}c@{}}Similarity\\ (virus)\end{tabular}}} & \multicolumn{1}{c}{\textbf{\begin{tabular}[c]{@{}c@{}}Word\\ (china)\end{tabular}}} & \multicolumn{1}{c}{\textbf{\begin{tabular}[c]{@{}c@{}}Similarity\\ (china)\end{tabular}}} & \multicolumn{1}{c}{\textbf{\begin{tabular}[c]{@{}c@{}}Word\\ (chinese)\end{tabular}}} & \multicolumn{1}{c}{\textbf{\begin{tabular}[c]{@{}c@{}}Similarity\\ (chinese)\end{tabular}}} & \multicolumn{1}{c}{\textbf{\begin{tabular}[c]{@{}c@{}}Word\\ (virus)\end{tabular}}} & \multicolumn{1}{c}{\textbf{\begin{tabular}[c]{@{}c@{}}Similarity\\ (virus)\end{tabular}}} \\ \midrule
xinjiang                                                                            & 0.735                                                                                     & quidpr                                                                                & 0.656                                                                                       & infect                                                                              & 0.787                                                                                      & ccp                                                                                 & 0.766                                                                                     & desensit                                                                              & 0.739                                                                                       & corona                                                                              & 0.761                                                                                     \\
turkic                                                                              & 0.702                                                                                     & taiwanese                                                                             & 0.652                                                                                       & antibodi                                                                            & 0.773                                                                                      & wuhan                                                                               & 0.759                                                                                     & chinavirus                                                                            & 0.735                                                                                       & viru                                                                                & 0.757                                                                                     \\
tibet                                                                               & 0.693                                                                                     & mainland                                                                              & 0.615                                                                                       & malnutrit                                                                           & 0.756                                                                                      & wuhanvirus                                                                          & 0.758                                                                                     & scapegoa                                                                              & 0.729                                                                                       & vir                                                                                 & 0.716                                                                                     \\
tradewar                                                                            & 0.666                                                                                     & cantonese                                                                             & 0.613                                                                                       & mutat                                                                               & 0.751                                                                                      & chinavirus                                                                          & 0.739                                                                                     & spokespeople                                                                          & 0.721                                                                                       & viruse                                                                              & 0.702                                                                                     \\
export                                                                              & 0.658                                                                                     & xijinp                                                                                & 0.609                                                                                       & measl                                                                               & 0.743                                                                                      & wuhancoronavius                                                                     & 0.732                                                                                     & communist                                                                             & 0.718                                                                                       & coronavirus                                                                         & 0.699                                                                                     \\
dcep                                                                                & 0.658                                                                                     & xinjiang                                                                              & 0.608                                                                                       & pathogen                                                                            & 0.738                                                                                      & prc                                                                                 & 0.731                                                                                     & ccp                                                                                   & 0.713                                                                                       & corana                                                                              & 0.698                                                                                     \\
xijinp                                                                              & 0.655                                                                                     & sabahan                                                                               & 0.602                                                                                       & outbreak                                                                            & 0.731                                                                                      & ccpvirus                                                                            & 0.726                                                                                     & disinf                                                                                & 0.712                                                                                       & covd                                                                                & 0.696                                                                                     \\
chongq                                                                              & 0.651                                                                                     & china                                                                                 & 0.602                                                                                       & inflamm                                                                             & 0.729                                                                                      & chinesevirus                                                                        & 0.726                                                                                     & wuhanvirus                                                                            & 0.711                                                                                       & caronavirus                                                                         & 0.690                                                                                     \\
sprat                                                                               & 0.642                                                                                     & uyghur                                                                                & 0.593                                                                                       & diseas                                                                              & 0.727                                                                                      & culpabl                                                                             & 0.722                                                                                     & incompe                                                                               & 0.705                                                                                       & carona                                                                              & 0.689                                                                                     \\
eros                                                                                & 0.638                                                                                     & danish                                                                                & 0.591                                                                                       & infecti                                                                             & 0.726                                                                                      & asshoe                                                                              & 0.714                                                                                     & vrisu                                                                                 & 0.700                                                                                       & coronaviri                                                                          & 0.689                                                                                     \\
whereshunt                                                                          & 0.630                                                                                     & dcep                                                                                  & 0.590                                                                                       & immune                                                                              & 0.720                                                                                      & silkroad                                                                            & 0.711                                                                                     & chicom                                                                                & 0.699                                                                                       & nipah                                                                               & 0.689                                                                                     \\
sichuan                                                                             & 0.630                                                                                     & tibetan                                                                               & 0.587                                                                                       & urinari                                                                             & 0.714                                                                                      & kne                                                                                 & 0.711                                                                                     & wuhancoronavius                                                                       & 0.698                                                                                       & desensit                                                                            & 0.687                                                                                     \\
eastturkistan                                                                       & 0.629                                                                                     & bytedance                                                                             & 0.584                                                                                       & hpv                                                                                 & 0.714                                                                                      & wuhanflu                                                                            & 0.705                                                                                     & reflexive                                                                             & 0.698                                                                                       & biochem                                                                             & 0.681                                                                                     \\
pork                                                                                & 0.627                                                                                     & cpec                                                                                  & 0.575                                                                                       & influenza                                                                           & 0.713                                                                                      & boycottchina                                                                        & 0.705                                                                                     & chinesevirus                                                                          & 0.697                                                                                       & syphili                                                                             & 0.678                                                                                     \\
muslim                                                                              & 0.623                                                                                     & counterpart                                                                           & 0.574                                                                                       & vaccine                                                                             & 0.712                                                                                      & madeinchina                                                                         & 0.704                                                                                     & wumao                                                                                 & 0.695                                                                                       & flue                                                                                & 0.676                                                                                     \\
munit                                                                               & 0.621                                                                                     & refut                                                                                 & 0.573                                                                                       & bacteria                                                                            & 0.708                                                                                      & communist                                                                           & 0.703                                                                                     & prc                                                                                   & 0.692                                                                                       & wuflu                                                                               & 0.675                                                                                     \\
mainland                                                                            & 0.621                                                                                     & cultu                                                                                 & 0.571                                                                                       & cardiovascular                                                                      & 0.708                                                                                      & chinali                                                                             & 0.701                                                                                     & wuhanflu                                                                              & 0.686                                                                                       & wuhancoronavius                                                                     & 0.673                                                                                     \\
shandong                                                                            & 0.619                                                                                     & tibet                                                                                 & 0.570                                                                                       & cannabinoid                                                                         & 0.707                                                                                      & chinazi                                                                             & 0.697                                                                                     & kungflu                                                                               & 0.683                                                                                       & distrac                                                                             & 0.672                                                                                     \\
cpec                                                                                & 0.614                                                                                     & wechat                                                                                & 0.560                                                                                       & vaccin                                                                              & 0.703                                                                                      & wumao                                                                               & 0.694                                                                                     & volunte                                                                               & 0.682                                                                                       & moronavirus                                                                         & 0.671                                                                                     \\
communist                                                                           & 0.610                                                                                     & warship                                                                               & 0.560                                                                                       & ebola                                                                               & 0.701                                                                                      & chinaisasshoe                                                                       & 0.693                                                                                     & china                                                                                 & 0.682                                                                                       & crono                                                                               & 0.669                                                                                     \\ \bottomrule
\end{tabular}%
}
\caption{Top 20 most similar words to the words ``china,'' ``chinese,'' and ``virus'' for the first and last trained word2vec models on Twitter.}
\label{tab:top_20_first_last_twitter}
\end{table*}

\begin{figure}[t!]
\centering
\includegraphics[width=\columnwidth]{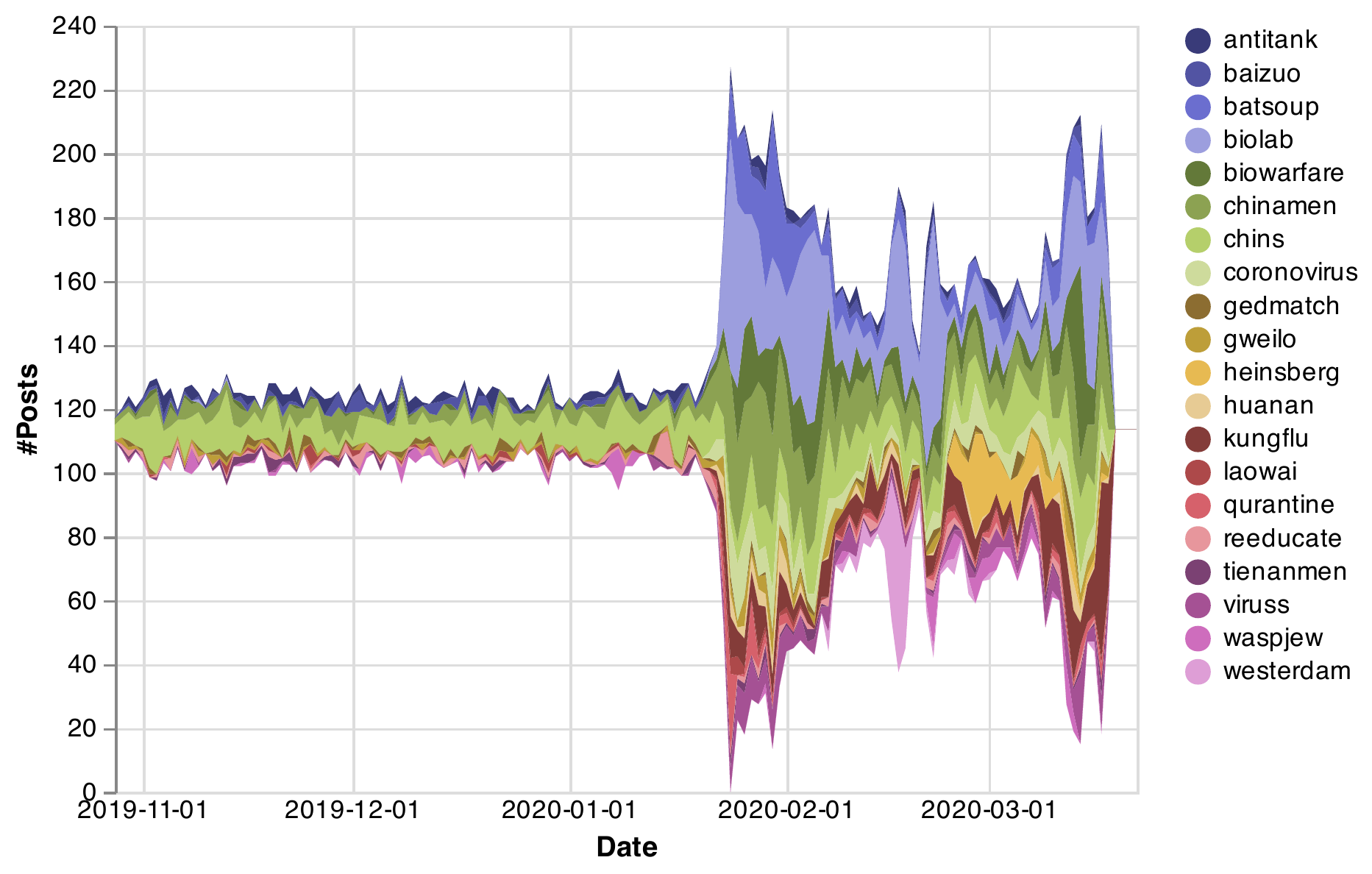}
\caption{Visualization of the emergence of new words related to ``chinese'' over time on 4chan's \dspol.}
\label{fig:new_terms_4chan}
\end{figure}

\descr{Discovering new terms.} Next, we aim to study how new terms, related to ``chinese,'' emerge on 4chan's \dspol and how their popularity changes over the course of our dataset.
To achieve this, we make use of the terms extracted from the vocabularies of the  trained word2vec models on 4chan's \dspol.
Specifically, we initially extract the vocabulary from the model trained on historical data (\wvcontrol{}) and treat it as our base vocabulary. 
Then, for each weekly trained model (\wvweekly{}), we extract the vocabulary and compare the terms with our base vocabulary: for each term that is new, we add it to our base vocabulary treat it as a \emph{new term}.
Since, we want to find new terms that are related to Chinese, we filter out all new terms that have a cosine similarity below 0.5 in the weekly trained model for which we discovered the new term.
Overall, using the above methodology, we manage to discover a total of 50 new terms.
Then, we visualize the popularity of the 20 most popular new terms of the course of our dataset in Figure~\ref{fig:new_terms_4chan}.

We observe the emergence of several interesting words during the the end of January, 2020.
First, we observe the emergence of terms like ``batsoup,'' likely indicating that \dspol users are discussing the fact that the COVID-19 outbreak, allegedly started by Chinese people consuming bats. 
Second, by the same time, we observe the emergence of ``biolab'' and ``biowarfare.'' 
The use of these words indicate that \dspol users discuss various conspiracy theories on how the COVID-19 virus was created on a lab or how it can be used as a bioweapon.
Interestingly, these terms are persistent from their emergence till the end of our datasets, indicating that these theories are generally appealing to 4chan's userbase.
Other interesting new terms include the terms ``kungflu,'' which an offensive term towards Chinese people related to the COVID-19 virus, and ``heinsberg,'' which is the center of the outbreak in Germany and indicates that \dspol users was discussing about it, especially during the end of February, 2020 and beginning of March, 2020.

The echo chamber effect \cite{zannettou2018gab} performs significantly on 4chan, that the narratives towards Covid-19 are consistently blaming China, and being racist, or spreading conspiracy theory, which alarms for the risk of information manipulation ~\cite{bond201261}~\cite{vosoughi2018spread}.
Previous studies on social networks have shown that a small number of zealots can distort collective decisions, especially on ambiguous events ~\cite{woolley2016automating}~\cite{stewart2019information}. 

\begin{figure*}[t!]
\center
\subfigure[$\mathcal{W}_{t=0}$]{\includegraphics[width=0.8\textwidth]{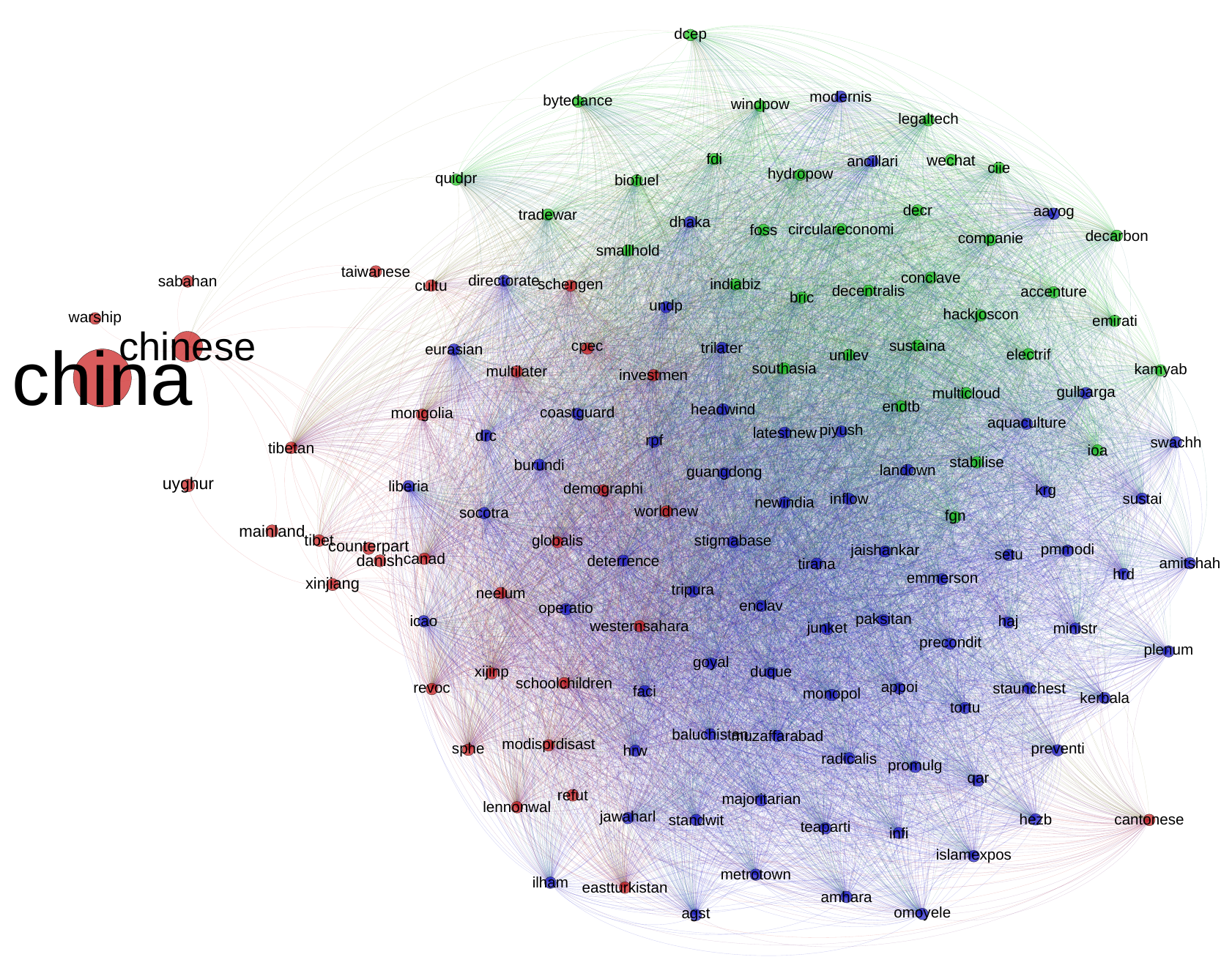}\label{fig:graph_chinese_twitter_first}}
\subfigure[$\mathcal{W}_{t=-1}$]{\includegraphics[width=0.8\textwidth]{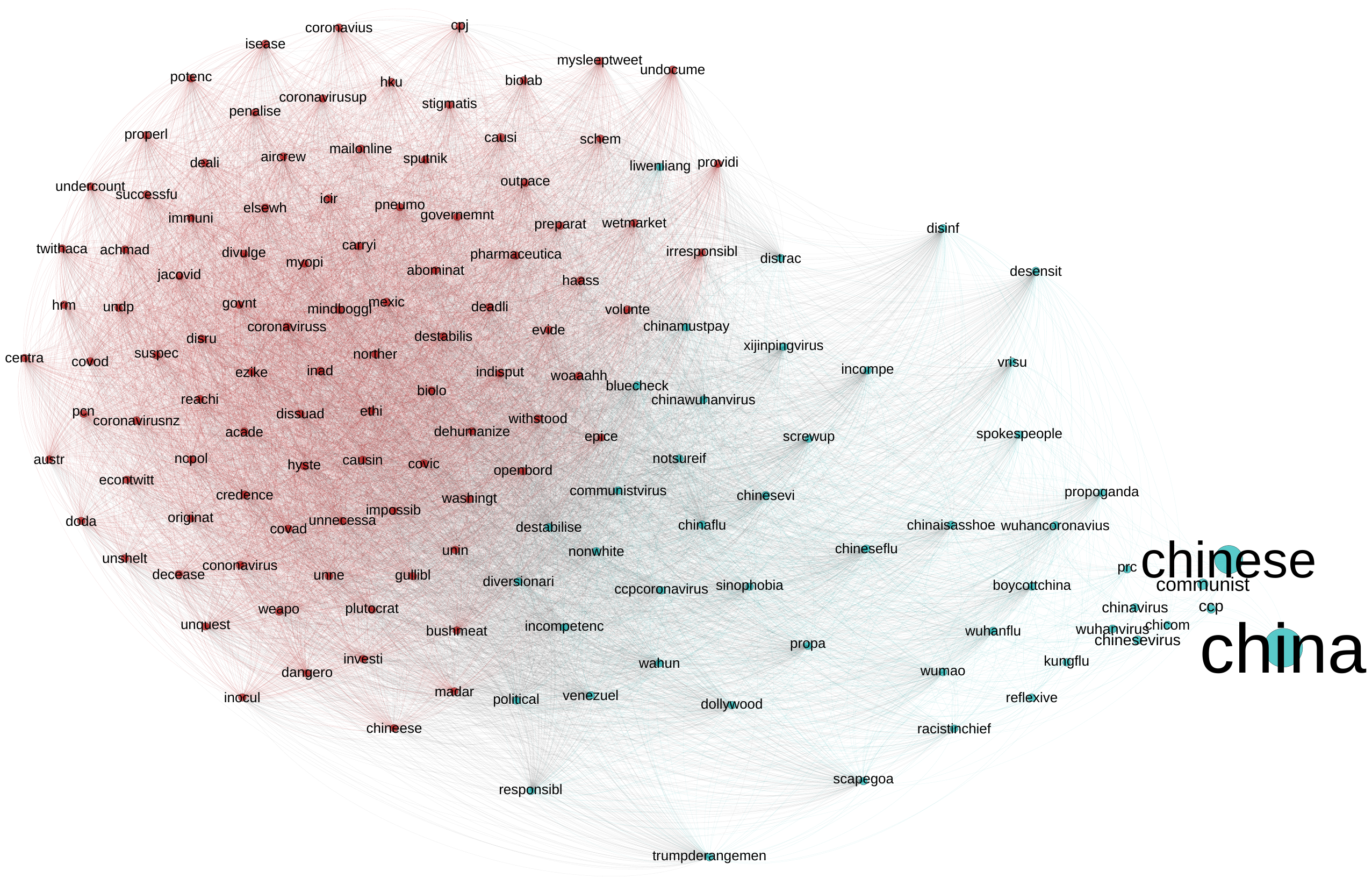}\label{fig:graph_chinese_twitter_last}}
\caption{Visualization of a 2-hop graph from the word ``chinese'' on Twitter using the first and last weekly word2vec models.} 
\label{fig:graph_chinese_twitter_overtime}
\end{figure*}

\subsection{Evolution on Twitter}
\label{section:evolve_twitter}

Now, we focus on the Sinophobic language evolution on Twitter.
We follow the same methodology used in Section~\ref{section:evolve_4chan}.
The corresponding results are depicted in Table~\ref{tab:top_20_first_last_twitter} and Figure~\ref{fig:graph_chinese_twitter_overtime}.

From Table~\ref{tab:top_20_first_last_twitter}, we can observe that during the first week covered by our Twitter dataset, many similar terms to ``china'' and ``chinese'' are related to politics, such as ``tradewar.''
This is again quite different from the result on \dspol (see Table~\ref{tab:top_20_first_last}).
Meanwhile, for ``virus,'' the most similar terms are also related to diseases.

However, when checking results on our last week Twitter data, we observe that many Sinophobic terms
appear to be semantically similar to ``china'' and ``chinese,'' such as ``chinazi.''
As in 4chan's \dspol (see Table~\ref{tab:top_20_first_last}), newly created Sinophobic terms, including ``chinavirus'' and ``kungflu,'' appear to be close in context as well.
For example, a Twitter user posted: \emph{``I agree. Too specific.  It's obviously called the kungflu.  It's kicking all of our asses regardless of denomination.''}

Moreover, many terms with similar contexts to ``china'' and ``chinese'' in our last week Twitter dataset are still about politics.
In contrast to the first week Twitter data, these political-related terms are related to COVID-19, e.g., ``ccpvirus,'' and some of these terms even convey the meaning of revenge and punishment towards China, such as ``boycottchina.''
For instance, a Twitter user posted: \emph{``\#ChineseVirus is chinesevirus. One name.  \#BoycottChina \#ChinaLiesPeopleDie.''}

\begin{figure*}[t!]
\center
\subfigure[]{\includegraphics[width=0.49\textwidth]{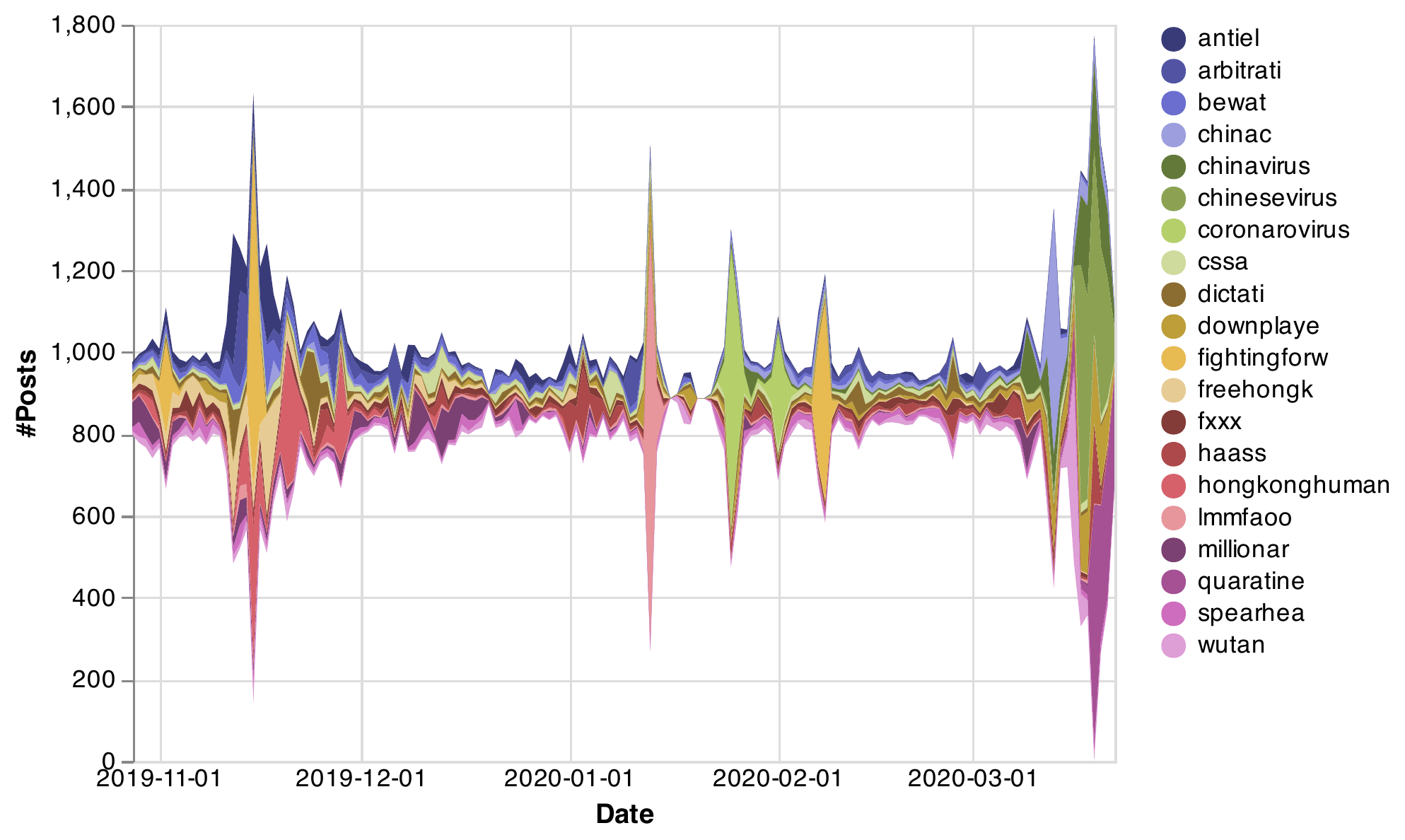}\label{fig:streamgraph_5}}
\subfigure[]{\includegraphics[width=0.49\textwidth]{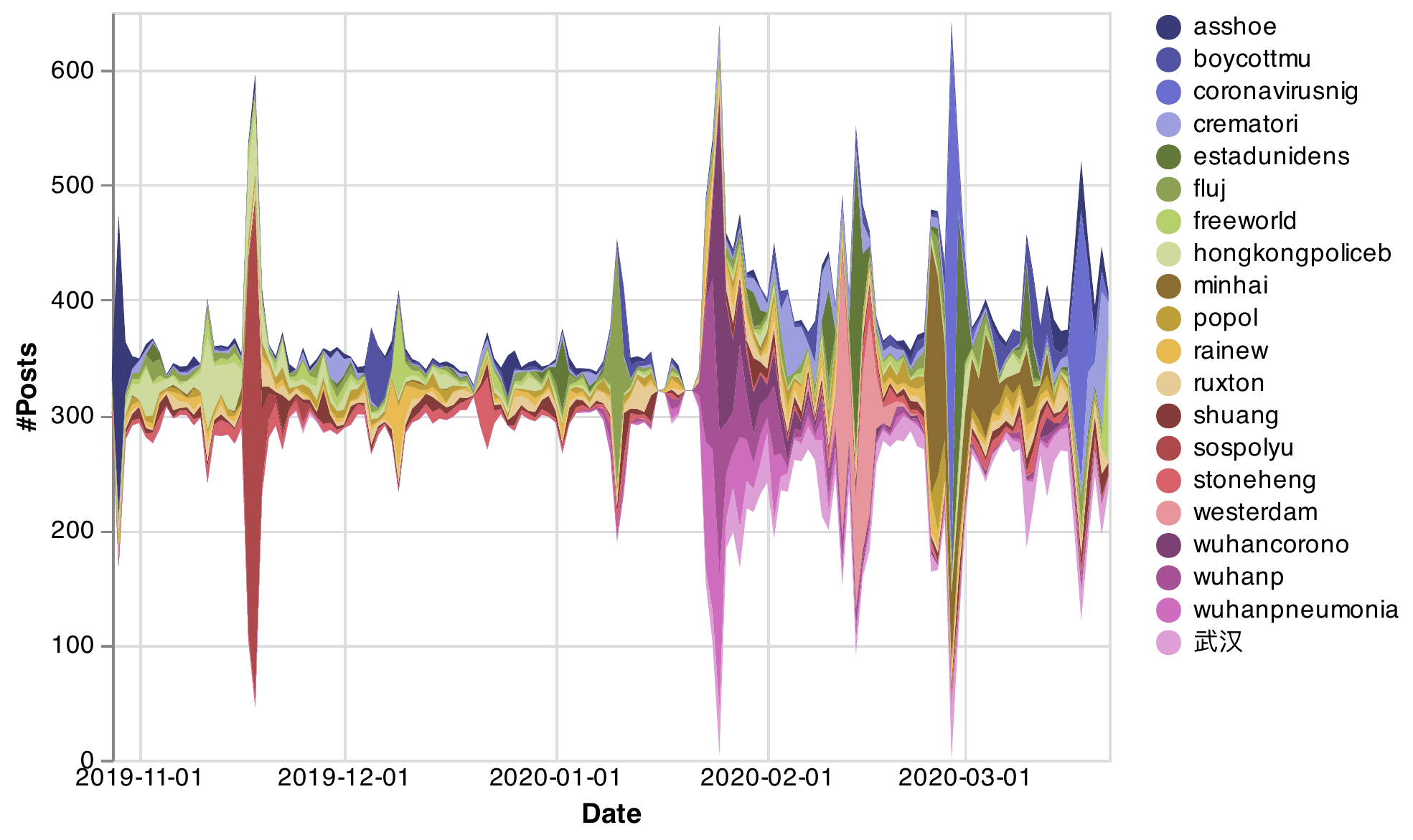}\label{fig:streamgraph_7}}\caption{Visualization of the emergence of new words related to ``chinese'' over time on Twitter.} 
\label{fig:new_terms_twitter}
\end{figure*}

\begin{figure*}[t!]
\center
\subfigure[chinese - virus]{\includegraphics[width=0.49\textwidth]{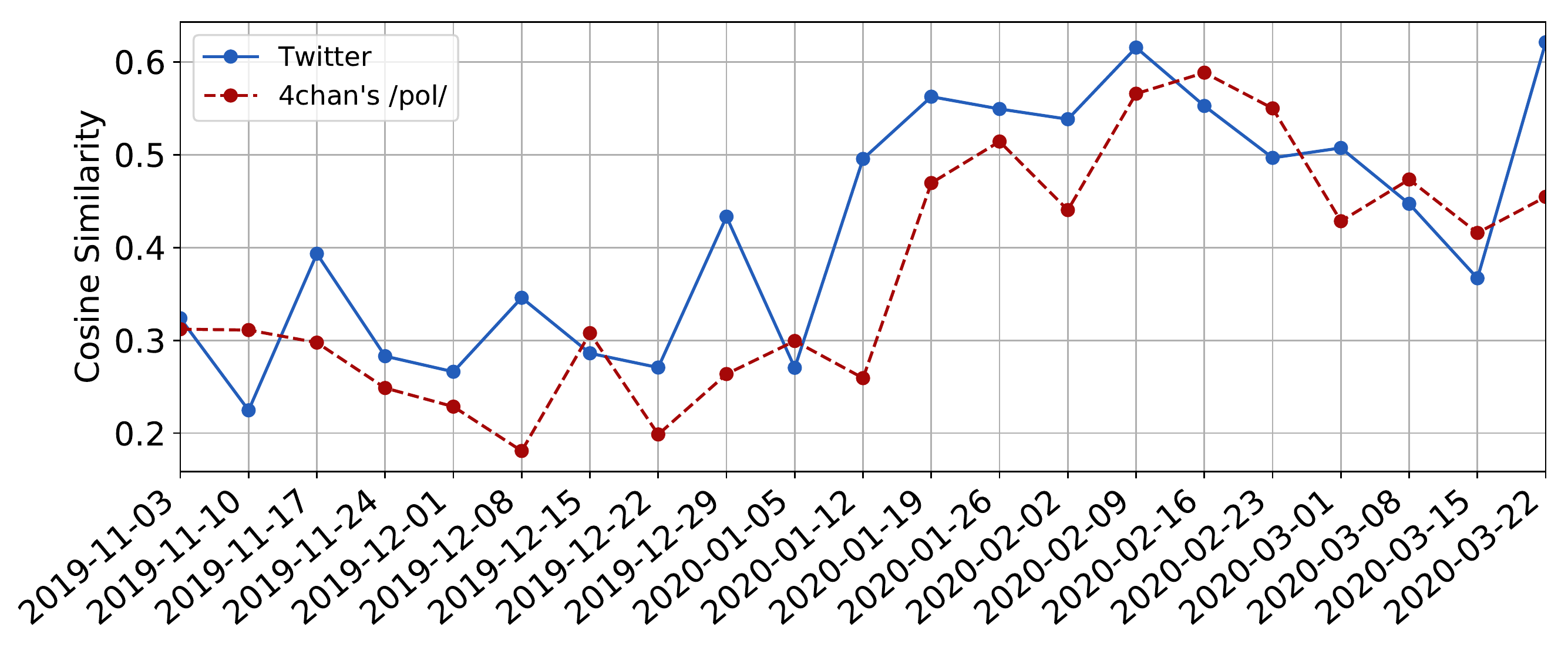}\label{fig:chinese_virus}}
\subfigure[chinese - chink]{\includegraphics[width=0.49\textwidth]{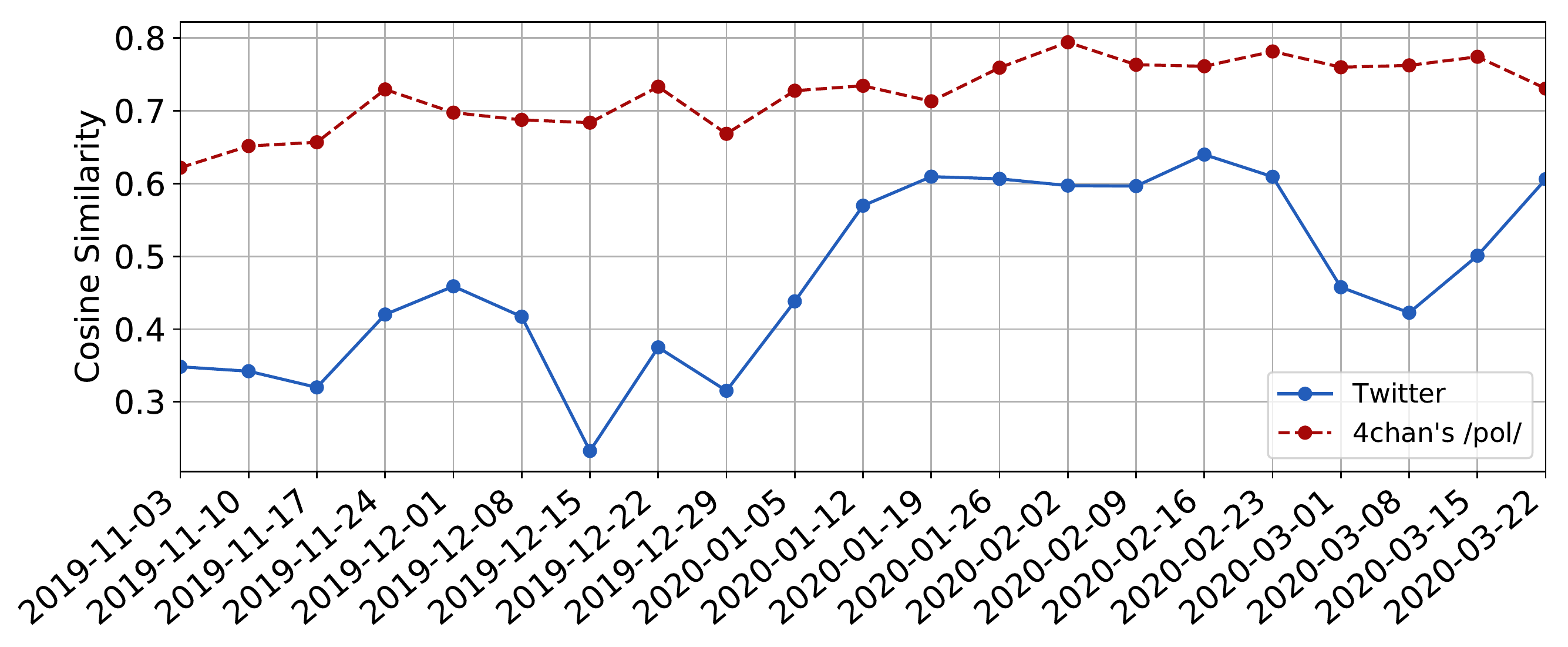}\label{fig:chinese_chink}}
\subfigure[chinese - bat]{\includegraphics[width=0.49\textwidth]{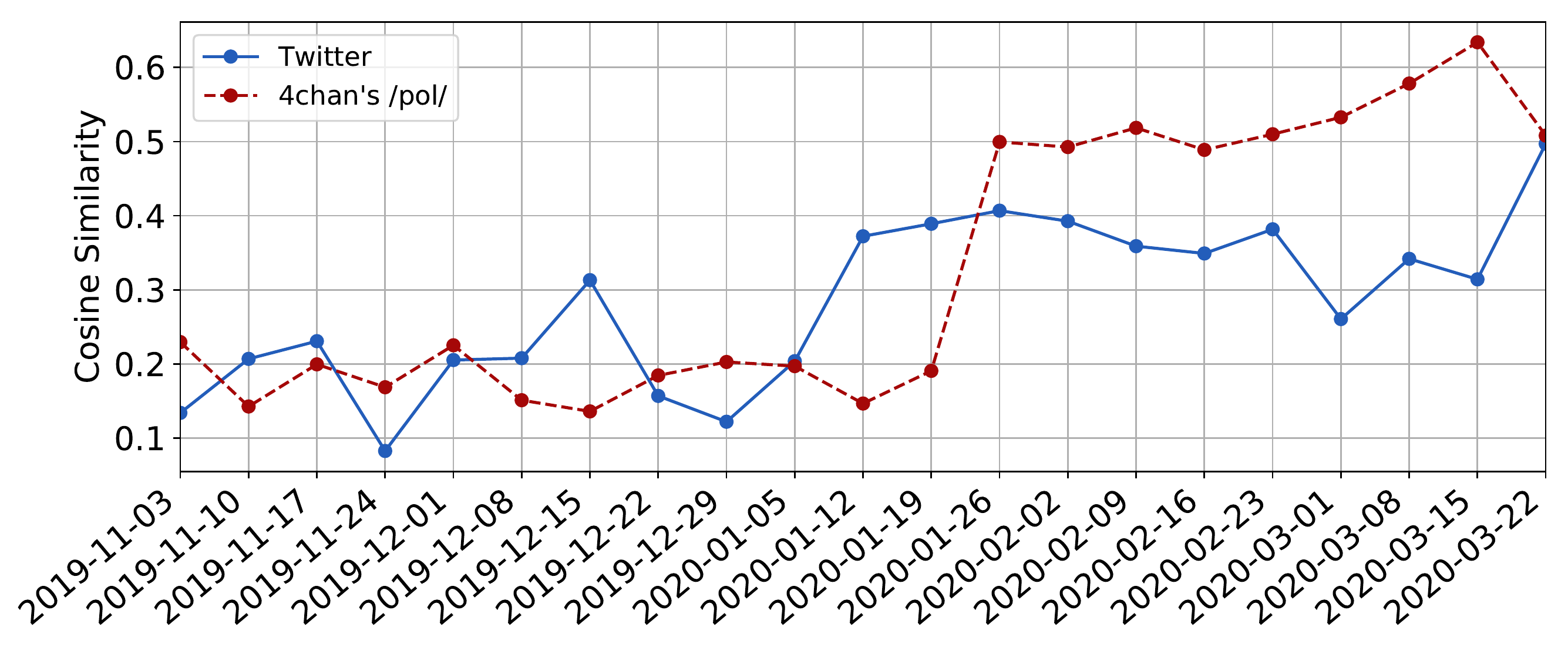}\label{fig:chinese_bat}}
\subfigure[chinese - pangolin]{\includegraphics[width=0.49\textwidth]{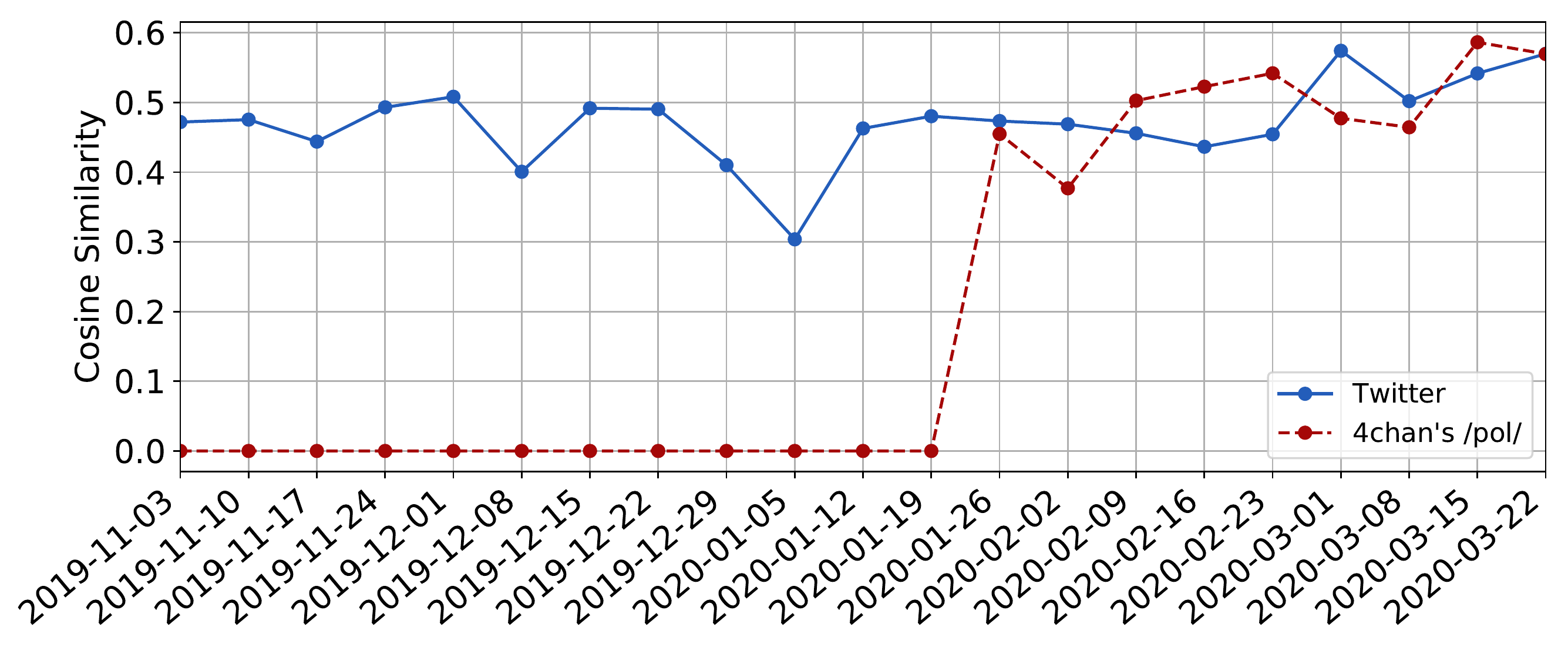}\label{fig:chinese_pangolin}}
\caption{Cosine similarities between various terms over time.} 
\label{fig:evolution_terms}
\end{figure*}

By looking into the graphs obtained from the first and last weekly trained word2vec models (see Figure~\ref{fig:graph_chinese_twitter_overtime}) we again observe substantial differences between the first and last models.
The graph from the first model includes mainly words related to China and other Asian regions, as well as words used for discussing matters related to China, e.g., ``tradewar.''
On the other hand, for the graph obtained from the last model, we observe several terms related to COVID-19 like ``chinavirus,'' ``chinesevirus,'' ``chineseflu,'' and ``chinaisasshoe.'' 
This indicates a shift towards the use of racist terms related to Chinese people after the COVID-19 outbreak on Twitter.
We also observe some terms that appear related to the behavior of Donald Trump.
For instance, the term ``racistinchief'' is likely related to the fact that Donald Trump calls the COVID-19 virus as ``Chinese Virus,'' and this was discussed on Twitter.
For instance, a Twitter user posted: \emph{``Trump’s a real asshole, just in case y’all forget \#TrumpPandemic \#TrumpVirus \#RacistInChief.''}

\descr{Discovering new terms.}
To discover new terms from Twitter, we follow the same methodology with \dspol, as documented in Section~\ref{section:evolve_4chan}.
Overall, we discover a total of 713 new terms between October 28, 2019 and March 22, 2020.
Figure~\ref{fig:new_terms_twitter} visualizes a sample of 40 of the new terms according to their popularity and cosine similarity with the term ``chinese.''
We observe a lot of new terms relating to the Hong Kong protests emerging during November 2019, such as ``freehongk'' and ``hongkongpoliceb.''
Also, after the outbreak of the COVID-19 pandemic, we observe the emergence of a wide variety of terms around the end of January 2020.
Some notable examples include terms like ``chinavirus,'' ``chinesevirus,'' ``wuhanpneumonia,'' ``wuhancorono,'' etc.
These findings highlight that during important real-world events, such as the COVID-19 pandemic, language evolves and new terms emerge on Web communities like Twitter.
At the same time, it is particularly worrisome that we observe the appearance of new terms that can be regarded as Sinophobic like ``chinesevirus,'' which can possibly lead to hate attacks in the real-world, and almost certainly harm international relations.

\subsection{Semantic Changes between Words}

As the last part of our analysis, we set out to assess how the semantic distance between words change over the course of our datasets.
To do this, we leverage the weekly trained word2vec models (\wvweekly{}): for each word2vec model, we extract the cosine similarity between two terms and then we plot their similarities over time.
This allow us to understand whether two terms are mapped closer to the multi-dimensional vector space over time, hence visualizing if two terms are used more in similar context over time.
We show some examples in Figure~\ref{fig:evolution_terms}: the terms are selected based on our previous analysis.

We observe several interesting changes in the similarities between terms over time.
Specifically, for the terms ``chinese'' and ``virus'' (see Figure~\ref{fig:chinese_virus}) we observe a substantial increase in cosine similarity between these two terms over time, especially after the week ending on January 19, 2020. 
The cosine similarity on both Twitter and \dspol was below 0.5 in the early models, while after January 19, 2020, it is mostly over 0.5, with the last model having a similarity over 0.6. 
This indicates that the terms ``chinese'' and ``virus'' are used in more similar ways over time on both Twitter and \dspol.

Another example are the terms ``chinese'' and ``chink'' (see Figure~\ref{fig:chinese_chink}).
We observe that for both Twitter and \dspol the similarity between these terms increases over the course of our datasets.
Interestingly, the increase in cosine similarity between these terms is \emph{larger} for Twitter, 
likely indicating that Twitter users are more affected by the COVID-19 with regards to sharing Sinophobic content, while on \dspol the difference is smaller which indicates that \dspol users were affected less by COVID-19 when it comes to sharing Sinophobic content.

Finally, we illustrate also the cosine similarity differences between the terms  ``chinese'' and ``bat''/``pangolin'' in Figure~\ref{fig:chinese_bat} and~\ref{fig:chinese_pangolin}, respectively.
For ``bat,'' we observe that the cosine similarity was low during our first models and it substantially increased after the week ending on January 26, 2020. 
This indicates that both Twitter and \dspol users have started discussing the fact that the virus allegedly originates from ``bats'' around that specific time frame and they continued doing so until the end of our datasets.
For ``pangolin,'' we observe some differences across the two Web communities: on \dspol the users were not discussing pangolins at all before January 26, 2020 and after that they started discussing them with a high cosine similarity to the term ``chinese'' (over 0.4).
On the other hand, on Twitter we observe that users were discussing pangolins even before the COVID-19 outbreak.

\section{Conclusion}

To combat the COVID-19 pandemic, many governments have implemented unprecedented measures like social distancing and even government enforced, large-sacle quarantines.
This has resulted in the Web becoming an even \emph{more} essential source of information, communication, socialization.
Unfortunately, the Web is also exploited for disseminating disturbing and harmful information, including conspiracy theories and hate speech targeting Chinese people.

Scapegoating is a basic psycho-social mechanism to deal with stress.
Building upon the well known in-group favoritism/out-group hostility phenomenon, racist ideology has a long history of scapegoating.
A common scapegoating theme has been to equate the targeted people with a disease, either figuratively or literally.
When threatened by events outside our control, it is only ``natural'' to seek for external blame.
In the case of COVID-19, the entire world is threatened, and there is a ``natural'' external actor to blame.

In this paper, we make a first attempt to understand Sinophobic language on the social Web related to COVID-19.
To this end, we collect two large-scale datasets from 4chan's \dspol and Twitter over a period of five months.
Our results show that COVID-19 has indeed come with a rise of Sinophobic content on both fringe Web communities like \dspol and mainstream ones like Twitter.
Relying on word embeddings, we also observe the semantic evolution of Sinophobic slurs.
Moreover, our study also shows that many new Sinophobic slurs are created as the crisis progresses.

Our study has several implications for both society and the research community focusing on understanding and mitigating emerging social phenomena on the Web.
First, we showed that the dissemination of hateful content, and in particular Sinophobic content, is a cross-platform phenomenon that incubates both on fringe Web communities as well as mainstream ones. 
This prompts the need to have a multi-platform point-of-view when studying such emerging social phenomena on the Web.
Second, we showed that Sinophobic behavior evolves substantially, especially after life changing events like the COVID-19 pandemic.
This highlights the need to develop new techniques and tools to understand these changes in behavior and work towards designing and deploying counter-measures with the goal to prevent or mitigate real-world violence stemming from these behaviors.

While the COVID-19 crisis does provide a unique opportunity to understand the evolution of hateful language, our study should be also be taken as a call for action.
The Web has enabled much of society to keep going, or at least to maintain social connections with other humans, but it has also allowed, and potentially \emph{encouraged} the proliferation of hateful language at a time where we can afford it the least.

\descr{Acknowledgments.} We thank the anonymous reviewers for their comments.
This work was supported by the NSF under Grant 1942610.

\small{
\bibliographystyle{abbrv}
\bibliography{sigproc}
}
\end{document}